\documentclass[12]{article}
\usepackage{amsfonts, amsmath, appendix, natbib, xcolor, graphicx}

\newcommand\beq[1]{ \begin{equation}\label{#1} }
\newcommand{\eeq}{ \end{equation} }
\newcommand{\beqno}{ \[ }
\newcommand{\eeqno}{ \] }
\newcommand\beqa[1]{ \begin{eqnarray} \label{#1} }
\newcommand{\eeqa}{ \end{eqnarray} }
\newcommand{\beqano}{ \begin{eqnarray*} }
\newcommand{\eeqano}{ \end{eqnarray*} }
\newcommand\equ[1]{{\rm (\ref{#1})}}
\newcommand\lng[1]{{#1}}

\newtheorem{remark}{Remark}

\title{Charged dust close to outer mean-motion resonances in the heliosphere}
\author{
Christoph Lhotka$^1$ \quad C\u at\u alin Gale\c s$^2$ \\
        \small ${}^1$Space Research Institute,
        \small Austrian Academy of Sciences\\
        \small Schmiedlstrasse 6, A-8042 Graz, \underline{Austria},
	\texttt{christoph.lhotka@oeaw.ac.at}
\and
        \small ${}^2$Faculty of Mathematics,
        \small 'Al. I. Cuza' University of Ia\c si\\
        \small Bd. Carol I, 11, 7000506 Ia\c si, \underline{Romania},
	\texttt{cgales@uaic.ro}
}
\begin{document}
\maketitle

\begin{abstract}
We investigate the dynamics of charged dust close to outer mean-motion
resonances with planet Jupiter. The importance of the interplanetary magnetic
field on the orbital evolution of dust is clearly demonstrated. New dynamical
phenomena are found that do not exist in the classical problem of uncharged
dust. We find changes in the orientation of the orbital planes of dust
particles, an increased amount of \lng{chaotic orbital motions}, sudden
'jumps' in the resonant argument, and a decrease in time of temporary capture
due to \lng{the} Lorentz force.  Variations in the orbital planes of dust grain orbits
are found to be related to the angle between the orbital angular momentum and
magnetic axes of the heliospheric field and the rotation rate of the Sun. 
These variations are bound using a
simplified model derived from the full dynamical problem using first order
averaging theory.  It is found that the interplanetary magnetic field does not
affect the capture \lng{process, that} is still dominated by the other
non-gravitational forces.  Our study is based on a dynamical model in the
framework of the inclined circular restricted three-body problem.  Additional
forces include solar radiation pressure, solar wind drag, the
Poynting-Robertson effect, and the influence of a Parker spiral type
interplanetary magnetic field model. The analytical estimates are derived on
the basis of Gauss' form of planetary equations of motion.  Numerical results
are obtained by simulations of dust grain orbits together with the system of
variational equations. Chaotic regions in phase space are revealed by means of
Fast Lyapunov Chaos Indicators.
\end{abstract}


\section{Introduction}

\begin{figure}
\begin{center}
\includegraphics[width=0.75\linewidth]{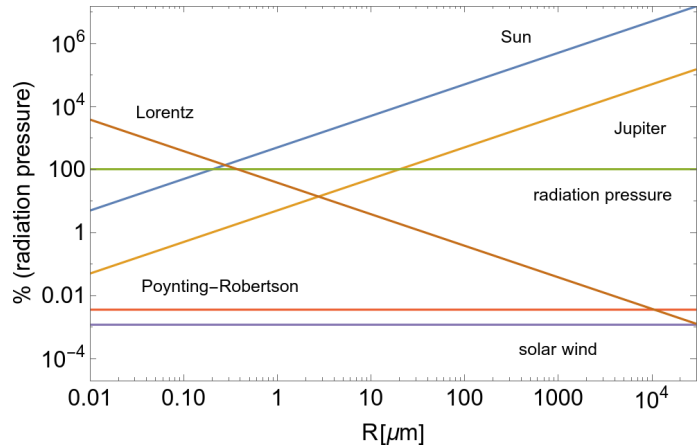}
\end{center}
\caption{Magnitudes of forces (evaluated using
\equ{e:Fg}, \equ{e:swpr}, \equ{e:FL2} at $8[AU]$ with dust grain density
$2.8[g/cm^3]$, and surface charge $5[V]$) in $\%$ of force due to solar
radiation pressure versus dust grain diameter $R$ (in microns).  We assume slow
solar wind conditions with mean solar wind speed equal $400 [km/s]$, and mean
background magnetic field strength defined at $1[AU]$ to be equal $3 [nT]$.}
\label{f:forze}
\end{figure}

Dust in space is subject to various non-gravitational forces originating from 
stellar radiation, i.e. due to the interaction of dust with
solar photons, the solar wind, as well as the heliospheric magnetic field. The
magnitudes of these forces do not only depend on the position and velocity of
these grains relative to the central star, but also on their physical and
chemical composition \citep[][]{kimman1998}. The orbital motion of dust is
clearly affected by the magnetic field in interplanetary space
\citep[][]{morgru1979, 1994A+A...286..915G}.  However, the role of it on motion close to mean
motion-resonances is still unclear.  Previous studies
\citep[][]{1993CeMDA..57..373S, 1994Icar..110..239B, 1994CeMDA..60..225B,
1997Icar..128..354L, 2003CeMDA..86..363J, 2008MNRAS.390.1491K,
2009A&A...501..367P, 2012MNRAS.422.1665K, 2014CeMDA.120...77P, 2015Icar..250..249L, 
2016MNRAS.460..524P} investigated the dynamical problem
under the assumption that dust is uncharged. This assumption is only valid in
the case where the dust grain is heavy enough and where the magnitude of the
Lorentz force is much smaller compared to the magnitudes of the other ones.
However, for sub-micron and micron sized particles this assumption is clearly
invalid as can already be seen in Fig.~\ref{f:forze}. The figure compares the
magnitudes of different forces that act on the orbit of micron-sized dust
grains in our solar system: gravitational attraction of the Sun and Jupiter,
solar radiation, force due to the Poynting-Robertson effect and solar wind drag, as
well as the Lorentz force.  We clearly see that Lorentz force starts as the
dominant force for submicron sized particles until radiation pressure, solar
attraction, and the force related to charge become similar in magnitude around
$R=0.2 [\mu m]$.  Electro-magnetic force decreases further for larger
particles until it becomes of the same order as Jupiter's gravity (around $3
[\mu m]$).  Interestingly enough, force related to charge is still larger in
magnitude than the other non-gravitational effects, that are usually taken into
account in these kinds of studies: solar wind drag and the force due to the
Poynting-Robertson effect.  For the choice of our parameters, Lorentz force
becomes smaller in magnitude than these other non-gravitational forces around
$1 [cm]$. The figure clearly demonstrates the importance of the interplanetary
magnetic field on charged-particle dynamics.  In recent years, the effect of
the interplanetary magnetic field on dust motion has already been included in
\cite{2006MNRAS.370.1876K}, where the authors studied the role of 
temperature-dependent optical properties on the life-time of dust particles.  In
\cite{2016ApJ...828...10L}, the authors investigated the role of the normal
component of the heliospheric magnetic field on the stability of dust motion in
interplanetary space. Both fore-mentioned studies excluded the phenomenon of 
resonant capture of dust. The charging history of dust grains is 
strongly influenced by the surrounding space plasma environment: currents due to the 
presence of free electrons and ions, secondary electron emission, and the photoelectric 
effect are known to be the dominant charging effects in interplanetary space 
\citep{2018P&SS..156...71P, 2014PhR...536....1M, 2011JPhD...44q4036P, 2006NPGeo..13..223P}. 
In studies of dusty plasmas the current balance 
equation allows to model the charge history of dust grains on time-scales relevant
for plasma physicists. Simulation studies show that charge of dust grains
becomes constant if the sum over all currents (from and to the surrounding 
space plasma) vanishes. The existence of stationary solutions enables us to
investigate long-term phenomena, i.e. the stability of dust grain orbits, in
interplanetary space with the assumption of constant charge. However,
it should be noted that all results based on this assumption
are only valid as long as we assume consistency of the surrounding space
plasma environment. This is the case in the current study. We assume a positive 
constant charge of dust grains solely due to the photoelectric effect. We thus neglect
variations in time of the charging process. For additional information, 
also see \cite{2002idpp.book.....S}.

The aim of the current study is to investigate the influence of the interplanetary
magnetic field on charged dust grain orbits close to outer mean-motion resonances.
Special focus is made to test the results in \citet{1994Icar..110..239B}
in presence of an additional Lorentz force term in the equations of motion of the
dust particles. Our study is based on a simple dynamical formulation of the problem,
i.e. the circular, inclined, restricted three-body problem. However, we also include
various non-gravitational effects, which are all detailed in Sec.~\ref{s:mod}.
For the interplanetary magnetic field we assume
the validity of a Parker spiral model \citep[][]{1958ApJ...128..664P,
2012bsw..book.....M} throughout the solar system, i.e. we neglect the effect
of solar cycle variations, solar flares, or coronal mass ejections. Already
this very simple formulation of the magnetic field in the heliosphere allows us
to demonstrate its effects on the orbits of charged dust grains: 1. A strong
influence on the inclination and ascending node of the orbital planes of the dust grains
with respect to the inertial plane, which is strongly related to the separation angle 
between the magnetic axis and the direction of the angular momentum. 2. A 'jumping' effect of the
resonant argument due to additional perturbations. 3. A decrease in capture time due to
the increased level of chaotic regions in phase space close to resonance.  \\

The model description can be found in Sec.~\ref{s:mod}, the isolated influence of the magnetic
field on the orientation of the orbital plane is described in Sec.~\ref{s:iso}. The detailed
numerical study of the unaveraged equations of motion is provided in Sec.~\ref{s:num}. The
main results are summarized and discussed in Sec.~\ref{s:sum}.

\section{Model and Methodology}
\label{s:mod}

We consider a  micrometer-sized dust particle orbiting in our solar system. We
study its orbital evolution by taking into account the gravity and the forces
due to the interaction of the dust grain with the solar radiation and the
interplanetary medium. We refer to the motion of the particle in a heliocentric,
ecliptic reference frame,
$\mathcal{G}_{Sun}=\{\vec{g}_x,\vec{g}_y,\vec{g}_z\}$, were $\vec g_{z}$ is
directed to the ecliptic pole, $\vec g_{x}$ is aligned with the direction to
the vernal equinox, and $\vec g_{y}$ completes the right handed set. The
equation of motion of the dust particle is given by:

\beq{e:dynsys}
m \ \ddot{\vec r} = \vec{F}_g + \vec{F}_{s/p} + \vec{F}_L \ .
\eeq

Here, $m$ and $\vec r$ are the mass and position of the particle in space and
$\ddot{\vec r}=d^2{\vec r}/dt^2$ is the acceleration vector. The particle
experiences a force $\vec{F}_g$ due to the gravity of the Sun and additional celestial
bodies in the solar system, the force $\vec{F}_{s/p}$ due to the interaction of the
particle with solar radiation and photons, and the Lorentz force $\vec{F}_L$ due to the
interplanetary magnetic field. In this study we aim to investigate a
simplified gravitational problem, where the particle only moves in the framework of
the restricted three-body problem \citep[][]{2000ssd..book.....M}:

\beq{e:Fg}
\vec F_g = -\frac{\mu m}{r^3}\vec r -\frac{\mu m_1 m}{m_0}\left(\frac{\vec r_1}{r_1^3} +
\frac{\vec r-\vec r_1}{\|\vec r-\vec r_1\|^3}\right) \ .
\eeq

Here, $\mu=Gm_0$ is the mass parameter that depends on the gravitational constant
$G$, the mass of the Sun $m_0$, while $m_1$ denotes the mass of the
main perturber, i.e. Jupiter in our case. In addition, the force depends on the
distance of the particle from the Sun $r$, as well as the position of Jupiter
${\vec r}_1$ and distance of the planet from the Sun $r_1$. Jupiter's orbit 
is assumed to be circular and included in the ecliptic plane. In the absence 
of $\vec{F}_{s/p}$,
$\vec{F}_L$, and the second term in \equ{e:Fg}, the dynamical problem given in
terms of \equ{e:dynsys} reduces to the integrable Kepler problem, while the
second term in \equ{e:Fg} renders \equ{e:dynsys} into a nearly integrable
problem, provided $m_1<<m_0$. \\

The form of $\vec{F}_{s/p}$ in \equ{e:dynsys} expanded up to first order
in $v/c$ is given by \citep[][equation 37]{2014Icar..232..249K}, and
equation 58 of \citet{2012MNRAS.421..943K}:

\beq{e:swpr}
\vec F_{s/p} =
-{\vec\nabla}\frac{\mu m \beta}{r}
-\frac{\mu m \beta}{r^2}\left(1+\frac{\eta}{Q}\right)
\left(\frac{\left(\dot{\vec r}\cdot\vec g_r\right)\vec g_r + \dot{\vec r}}{c}\right) \ .
\eeq

Here, dimensionless parameter $0\leq\beta\leq1$ denotes the ratio between the
magnitudes of the forces due to radiation pressure of the Sun and solar
gravity, parameter $\eta=1/3$ is the ratio of the magnitudes of the
forces due to the solar wind and the Poynting-Robertson effect, parameter $Q$
is the spectrally averaged efficiency factor (see
\citet{2012MNRAS.421..943K} for choice of parameters and further information on
the symbols). Parameter $c$ denotes the speed of light. The force also
depends on the unit vector $\vec{g}_r=\vec{r}/r$, which is directed towards the
position of the particle.  The isolated
force ${\vec F}_{s/p}$ alone is a velocity dependent force that cannot be
derived from a potential, and renders \equ{e:dynsys} to be a weakly dissipative
dynamical system \citep[see, e.g.][]{2012RCD....17..273C, 2013IJBC...2350036L}.
However, we note that the first term in \equ{e:swpr}, usually referred to as
radiation pressure, can be derived from the gradient of a potential. The
mathematical form of the potential coincides with the Kepler
problem. For this reason it is convenient to rewrite the equation of motion
\eqref{e:dynsys} in the form
\beq{e:dynsys2}
m \ \ddot{\vec r} = {\vec F}_{Kep} + {\vec F}_{Jup}+ \vec{F}_{s/p}' + \vec{F}_L \ ,
\eeq
where
\beq{e:FKep}
{\vec F}_{Kep} = -\frac{\mu(1-\beta) m}{r^3}\vec r \ ,
\eeq
while ${\vec F}_{Jup}$ and ${\vec F}_{s/p}'$ denote the second terms in
\equ{e:Fg} and \equ{e:swpr}, respectively. Clearly, ${\vec F}_{Jup}$ represents
Jupiter's gravitational attraction and the velocity--dependent component
${\vec F}_{s/p}'$ is due to the Poynting-Robertson effect and solar wind drag. From a
mathematical point of view, the net effect of the solar radiation pressure term alone is 
a reduction of the mass of the Sun by the factor $1-\beta$.

\begin{figure}
\begin{center}
\includegraphics[width=0.65\linewidth]{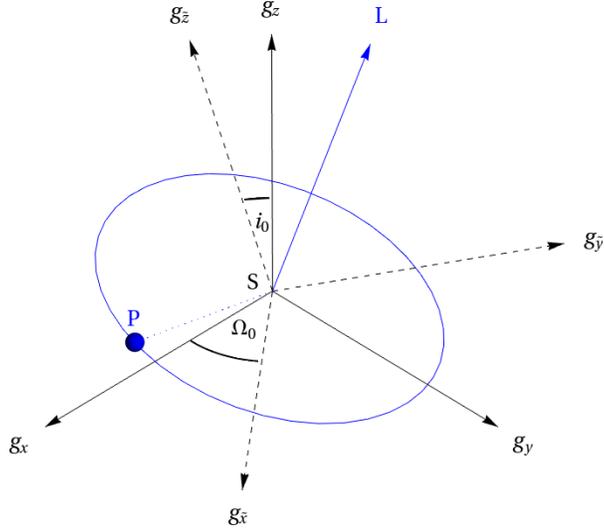}
\end{center}
\caption{Heliocentric reference frames: ecliptic $(\vec{g}_x,\vec{g}_y, \vec{g}_z)$ - thick,
equatorial $(\vec{g}_{\tilde x}, \vec{g}_{\tilde y}, \vec{g}_{\tilde z})$ - dashed. The orbit 
frame is shown in blue with vector of angular momentum $\vec L$.}
\label{f:graph}
\end{figure}

The last term that appears in \equ{e:dynsys} is the Lorentz force $\vec{F}_L$ that is given
in terms of \citep[][equation 10]{1994A+A...286..915G}:

\beq{e:FL2}
{\vec F}_L = q \left({\dot{\vec r}} - {\vec u}_{sw}\right)\times\vec B \ ,
\eeq
where $q$ is the charge of the dust particle, ${\vec u}_{sw}$
denotes the velocity vector of a radially expanding and uniform solar wind, that is
 \beq{e:solarwind}
{\vec u}_{sw} = u_{sw} \vec {g}_{r}\ ,
\eeq
 and $\vec B=\vec{B}(\vec r)$ is the interplanetary magnetic field.  We make
use of the so-called Parker spiral model, which is given in
\citet{2010JGRA..11510112W}. In order to provide the mathematical expression of
the magnetic field $\vec B$, we introduce first the following additional
reference frames as shown in Fig.~\ref{f:graph}.

The heliocentric reference frame $\mathcal{G}_{Sun}=\{\vec{g}_x, \vec{g}_y,
\vec{g}_z\}$ described above, we consider the heliocentric frames $
\tilde{\mathcal{G}}_{Sun}=\{\vec{g}_{\tilde x}, \vec{g}_{\tilde y},
\vec{g}_{\tilde z}\}$ and $ \tilde{\mathcal{G}}^{sc}_{Sun}=\{\vec{g}_{r},
\vec{g}_{\theta}, \vec{g}_{\phi}\}$ defined as follows: The unit vectors ($\vec
g_{\tilde x}$, $\vec g_{\tilde y}$) span the Sun's equatorial plane, with $\vec
g_{\tilde x}$ directed along the line of intersection between the ecliptic and
equatorial planes, and $\vec g_{\tilde z}$ is aligned with the rotation axis of
the Sun. Then, the following relation holds
\beqa{e:ec2eq}
\vec g_{\tilde x} &=&
\cos\left(\Omega_0\right)\vec g_{x} +
\sin\left(\Omega_0\right)\vec g_{y} \ , \nonumber \\
\vec g_{\tilde y} &=&
\cos\left(i_0\right)\left[
-\sin\left(\Omega_0\right)\vec g_{x} +
\cos\left(\Omega_0\right)\vec g_{y}\right] +
\sin\left(i_0\right) \vec g_{ z} \ , \nonumber \\
\vec g_{\tilde z} &=&
\sin\left(i_0\right)\left[
\sin\left(\Omega_0\right)\vec g_{x} -
\cos\left(\Omega_0\right)\vec g_{y}\right] +
\cos\left(i_0\right) \vec g_{z} \ ,
\eeqa
where $\Omega_0$ is the angle in the ecliptic plane between the direction of the vernal equinox and the line of nodes
between the equatorial and ecliptic planes, and $i_0$ is the
inclination of the Sun's equator with respect to the ecliptic plane.

Let $(r,\theta ,\phi )$ be the spherical coordinates in the equatorial system
of reference, that is $r \geq 0$ is the radial distance, $0 \leq  \theta \leq
\pi $ is the polar angle and $0\leq \phi <2\pi$ is the azimuthal angle.
$\vec{g}_{r}$, $\vec{g}_{\theta}$, $\vec{g}_{\phi}$ are the local orthogonal
unit vectors in the directions of increasing $r$, $\theta$ and $\phi$,
respectively, defined as
\beqa{e:s2e}
\vec g_{r} &=&
\sin\left(\theta\right)\left[
\cos\left(\phi\right)\vec g_{\tilde x} +
\sin\left(\phi\right)\vec g_{\tilde y}\right] +
\cos\left(\theta\right) \vec g_{\tilde z} \ , \nonumber \\
\vec g_{\theta} &=&
\cos\left(\theta\right)\left[
\cos\left(\phi\right)\vec g_{\tilde x} +
\sin\left(\phi\right)\vec g_{\tilde y}\right] -
\sin\left(\theta\right) \vec g_{\tilde z} \ , \nonumber \\
\vec g_{\phi} &=&
-\sin\left(\phi\right) \vec g_{\tilde x} + \cos\left(\phi\right)\vec g_{\tilde y} \ .
\eeqa

For later convenience, we also introduce a reference frame centered at the
position of the dust grain, denoted by $\mathcal{G}^{RTN}_{Dust}=\{\vec{g}_R,
\vec{g}_T, \vec{g}_N\}$. This orbital frame of reference is defined as follows.
$\vec g_{R}=\vec g_{r}$ is directed along the radial vector that connects the
origin (located in the center of the Sun) and the position of the dust grain.
The unit vector $\vec g_{N}$ is aligned with the direction of the angular
orbital momentum of the unperturbed dust grain, i.e. subject only to the
gravitational attraction of the Sun. The right handed set is completed by means
of the unit vector $\vec g_{T}$ oriented in the direction of the velocity
vector of the grain. We note that this $RTN$ system is different from the
$RTN$ system that has been defined in literature to implement the
interplanetary magnetic field
\citep[see][]{10.1111/j.1365-2966.2006.10612.x}. In this study, the
electromagnetic fields are solely referred in a $(r,\theta,\phi)$
system.

Let the angles $i$, $\omega$, $\Omega$, $f$ denote the inclination,
argument of perihelion, longitude of the ascending node, and true
anomaly of the dust grain orbit, defined with respect to the ecliptic system, respectively.
We define the true longitude $u=\omega+f$ and we introduce the
matrix
\beq{e:rm}
{\mathbf R} =
{\mathbf R}_3\left(-u\right)
{\mathbf R}_1\left(-i\right)
{\mathbf R}_3\left(-\Omega\right) \ ,
\eeq
where ${\mathbf R}_j(\psi)$ denotes
the rotation matrix\footnote{We note that the rotation matrices are set-up 
to use the vector-orientated convention and to give a matrix $\mathbf R$ so that 
$\mathbf R\cdot\vec r$ gives the rotated version of a vector 
$\vec r$ \citep[][]{Mathematica}.}
of angle $\psi$ around the axis $j$ with $j=1,3$, such that:
\beqano
{\mathbf R}_1\left(\psi\right)&=&\left(
\begin{array}{ccc}
 1 & 0 & 0 \\
 0 & \cos (\psi ) & -\sin (\psi ) \\
 0 & \sin (\psi ) & \cos (\psi ) \\
\end{array}
\right) \ , \nonumber \\
\nonumber \\
{\mathbf R}_3\left(\psi\right)&=&\left(
\begin{array}{ccc}
 \cos (\psi ) & -\sin (\psi ) & 0 \\
 \sin (\psi ) & \cos (\psi ) & 0 \\
 0 & 0 & 1 \\
\end{array}
\right)
\eeqano

Then, the components $F_x$, $F_y$, $F_z$  of the force $\vec F$, expressed in the reference frame $\mathcal{G}_{Sun}$, are transformed into
the components $F_R$, $F_T$, $F_N$ in the orbital RTN system $\mathcal{G}^{RTN}_{Dust}=\{\vec{g}_R, \vec{g}_T, \vec{g}_N\}$ through the relation:

\beq{e:o2e}
\left(F_R, F_T, F_N\right)^T = {\mathbf R}\left(F_x,F_y,F_z\right)^T \ .
\eeq

\paragraph{Parker spiral model.}
In this paper, the interplanetary magnetic field is described by the Parker
spiral model \citep[][]{2010JGRA..11510112W, 1987ApJ...315..700B}, given as:

\begin{figure}
\begin{center}
\includegraphics[width=0.65\linewidth]{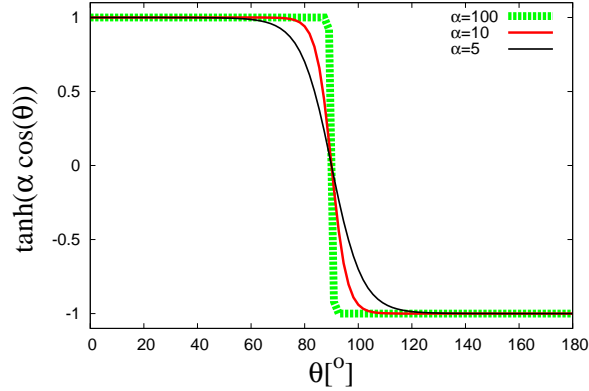}
\end{center}
\caption{The function $\tanh(\alpha \cos(\theta))$ for various values of the parameter $\alpha$.}
\label{f:Tanh}
\end{figure}

\beq{e:Bfld}
\vec B = \frac{A f\left(\theta\right)}{r^2}
\left(\vec g_r - \frac{\Omega_s r \sin\left(\theta\right)}{u_{sw}}\vec g_\phi\right) \ ,
\eeq
where the dimensional constant $A=B_0r_0^2$ is expressed in terms of the background
magnetic field strength\footnote{Here, we assume $B_0$ as a constant parameter. Subsequent works will include solar cycle variations, whose effects render periodical sign changes of $B_0$.} $B_0$, given at the reference distance $r_0$. The Lorentz force depends on the magnitude of the
radial component of the solar wind speed $u_{sw}$, the solar rotation rate $\Omega_s$, and the sign change function $f(\theta)$, which defines
the polarity of the magnetic field (magnetic north) of the Sun and can be given in terms of the Heaviside step function as $f(\theta)=1-2H(\theta-\pi/2)$. Thus, $f(\theta)$ is equal to $1$
above the equator ($\theta<\pi/2$) and $f(\theta)$  will change sign to $-1$ for $\theta>\pi/2$.
We note that $f(\theta)$ can easily be modified to incorporate additional effects,
like the sectorial structure of the interplanetary magnetic field.

Since $f(\theta)$ is discontinuous at $\theta=90^o$, we will replace this
function by the smooth function $\tanh(\alpha \cos(\theta))$, where $\alpha$ is
a positive parameter. Figure~\ref{f:Tanh} shows that in simulations it is
enough to consider $\alpha \geq 100$.

Denoting $\tilde{x}$, $\tilde{y}$ and $\tilde{z}$ the components of $\vec{r}$ in the frame $\tilde{\mathcal{G}}^{sc}_{Sun}$, that is we have $\tilde{x}= r \sin \theta \cos \phi$, $\tilde{y}= r \sin \theta \sin \phi$, $\tilde{z}= r \cos \theta$, it is easy to write the term $r \sin \theta \, \,  \vec g_\phi$, appearing in \eqref{e:Bfld}, in the form
\beq{e:sinthetagphi}
r \sin \theta\, {\vec g}_{\phi} = -\tilde{y} {\vec g}_{\phi} + \tilde {x} {\vec g}_{\phi}= {\vec g}_{\tilde{z}} \times {\vec r} \ .
\eeq
Moreover, from the first equation of $\eqref{e:s2e}$ we obtain
\beq{e:costheta}
\cos \theta= \frac{{\vec r} \cdot {\vec g}_{\tilde{z}}}{r} \ .
\eeq
Therefore, by using the relations \eqref{e:sinthetagphi} and \eqref{e:costheta}, the relation \eqref{e:Bfld} can be rewritten in the form
\beq{e:Bfldbis}
\vec B = \frac{B_0 r_0^2}{r^2}
\left(\frac{\vec r}{r} - \frac{\Omega_s} {u_{sw}} {\vec g}_{\tilde{z}} \times {\vec r} \right) \tanh\Bigl(\alpha \frac{{\vec r} \cdot {\vec g}_{\tilde z}}{r} \Bigr) \ .
\eeq

Combining \eqref{e:FL2}, \eqref{e:solarwind} and  \eqref{e:Bfldbis}, we obtain that the Lorentz force ${\vec F}_L$ has the form:

\beq{e:cLF}
{\vec F}_L=-\frac{q B_0 r_0^2}{r^2} \Bigl[{\frac{1}{r} \,\vec r} \times \dot{\vec r} +\frac{\Omega_s}{r} {\vec r} \times ( {\vec r} \times {\vec g}_{\tilde z}) +\frac{\Omega_s}{u_{sw}} ({\vec r} \times  {\vec g}_{\tilde z}) \times \dot{\vec r} \Bigr] \tanh\Bigl(\alpha \frac{{\vec r} \cdot {\vec g}_{\tilde z}}{r} \Bigr)
\eeq
where ${\vec g}_{\tilde z}$ is given by the third relation of $\eqref{e:ec2eq}$.

\begin{remark}
Lorentz force defined by \eqref{e:cLF} is conservative, that is there exists the function
\beq{e:Upsilon}
\Upsilon (\vec r) = -\frac{q B_0 r_0^2 \Omega_s}{\alpha} \ln \Bigr[\cosh\Bigl(\alpha \frac{{\vec r} \cdot {\vec g}_{\tilde z}}{r} \Bigr) \Bigr]\ ,
\eeq
such that
$$ {\vec F}_L \cdot d {\vec r}=-d \Upsilon. $$
\end{remark}
Indeed, from \eqref{e:cLF} we deduce
$${\vec F}_L \cdot d {\vec r}= -\frac{q B_0 r_0^2 \Omega_s}{r^3} \Bigl[ ({\vec r} \cdot {\vec g}_{\tilde z}) \, ( {\vec r} \cdot d {\vec r} ) - r^2 ( {\vec g}_{\tilde z} \cdot d {\vec r}) \Bigr] \tanh\Bigl(\alpha \frac{{\vec r} \cdot {\vec g}_{\tilde z}}{r} \Bigr)$$
$$=q B_0 r_0^2 \Omega_s  \tanh\Bigl(\alpha \frac{{\vec r} \cdot {\vec g}_{\tilde z}}{r} \Bigr)\,  d\Bigl(\frac{{\vec r} \cdot {\vec g}_{\tilde z}}{r}\Bigr)=-d \Upsilon.$$

\begin{figure}
\begin{center}
\includegraphics[width=0.65\linewidth]{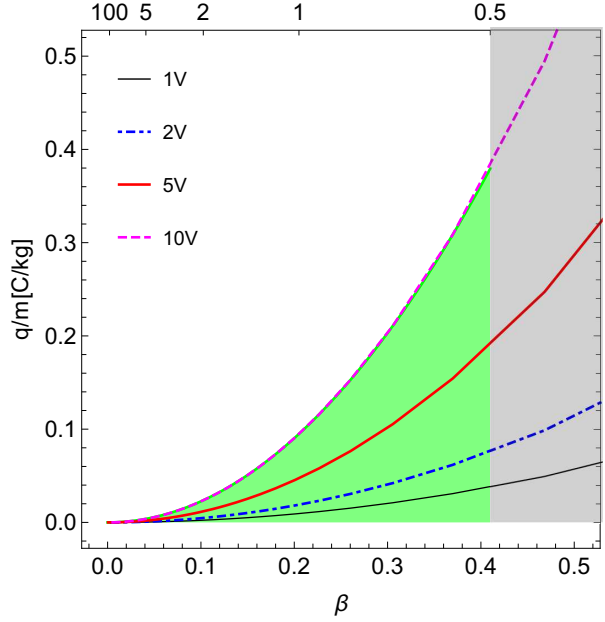}
\end{center}
\caption{Region of interest (light green) in the space $(\beta, \gamma)$.
Numbers at the top correspond to $R$ in $\mu m$.}
\label{f:pars}
\end{figure}

\paragraph{Choice of parameters.}
As described above, equation of motion \eqref{e:dynsys}, i.e.
\equ{e:Fg}--\equ{e:solarwind}, depends on the following parameters: the
ratio $q/m$ between the charge and mass of the dust grain, the ratio $\beta$
between the magnitudes of the forces due to solar radiation pressure and
the gravitational attraction of the Sun,  the ratio $\eta$ of the
magnitudes of the forces due to the solar wind and the Poynting-Robertson
effect, the spectrally averaged efficiency factor $Q$ and the solar wind speed
$u_{sw}$. Following \citet{2016ApJ...828...10L, 2006MNRAS.370.1876K}, in this 
paper we consider the following parameters as fixed constants: $\eta=1/3$, 
$Q=1$ and $u_{sw}=400\, [km/s]$.

Parameters $\beta$ and $q/m$ are related to other physical parameters of the
problem. Let the density and radius of the dust grain be parameterized by
$\rho$, and $R$, respectively. We assume that the dust grain is of spherical
shape, and has constant density and radius. Under this assumption the mass of
the grain is simply given by the formula $m=4\pi/3\rho R^3$.  On the other
hand, the charge $q$ of the dust grain is related to $U$, the surface electric
potential of the grain, by the formula: $q=4\pi\varepsilon_0\, U\,R$, where
$\varepsilon_0$ denotes the dielectric constant in vacuum.  Assuming perfectly
absorbing dust particles \citep[][]{1994Icar..110..239B} and setting
$\rho=2.8\, [g/cm^{3}]$, we obtain approximate formulas that are sufficient
for the purpose of our study:

\beq{e:pars}
\beta=0.205/R \quad \gamma=0.0094U/R^2 \ .
\eeq
Here,  $\gamma$ is the numerical value of $q/m$ expressed in $[C/kg]$ ($q/m=\gamma [C/kg]$), the particle radius $R$ is given in microns and the surface potential $U$ in Volts. Thus, the dimensionless quantities  $(\beta,\gamma)$ are given in terms of $(R, U)$ (and vice versa).
Table~\ref{t:pars} gives various quantities and numerical values that we use in our study.

\begin{table}
\caption{Parameters that we use in our study. $^{(a)}$ taken from NASA
planetary fact sheet {https://nssdc.gsfc.nasa.gov/planetary/factsheet/}. $^{(b)}$ calculated
from $\rho$, $U$, $R$. $^{(c)}$ obtained from $i_0$, $\Omega_0$.}
\label{t:pars}
\begin{tabular}{lll}
\hline\noalign{\smallskip}
symbol & values & reference \\
\noalign{\smallskip}\hline\noalign{\smallskip}
$a_1$ & $a_{J}$  & $^{(a)}$ \\
$\alpha$ & $100$ & this work \\
$B_0$ & 3nT & \cite{2012bsw..book.....M} \\
$\beta$ & 0\dots0.5 & \cite{1994Icar..110..239B} \\
$c$ & 299792458km/s & \cite{constBOOK} \\
$\varepsilon_0$ & $8.854187 \times 10^{-12}F/m$ & \cite{constBOOK} \\
$\eta$ & $1/3$  & \cite{2014MNRAS.443..213K} \\
$\gamma$& $0\dots0.5C/kg$ & $^{(b)}$ \\
$i_0$ & $7.15^o$ & \cite{2005ApJ...621L.153B} \\
$m_0$ & $M_\odot$ & $^{(a)}$  \\
$m$ & $4\pi/3\rho R^3$  & \cite{constBOOK} \\
$m_1$ & $m_J$ & $^{(a)}$ \\
$r_0$ & $1AU$ & \cite{2012bsw..book.....M} \\
$r_1$ & $a_1$ & $^{(a)}$ \\
$\Omega_0$ & $73.5^o$ & \cite{2005ApJ...621L.153B} \\
$\Omega_s^{-1}$ & $24.47d$ & \cite{2012bsw..book.....M} \\
$Q$ & $1$ & \cite{1994Icar..110..239B} \\
$R$ & $0\dots500\mu m$ & \cite{1994A+A...286..915G} \\
$\rho$ & $2.8g/cm^3$ & \cite{1994Icar..110..239B} \\
$u_{sw}$ &400km/s& \cite{2012bsw..book.....M}  \\
$U$ & $0\dots10V$  & \cite{2014PhR...536....1M} \\
$x_0,y_0,z_0$ & $(0.12,0.04,0.99)$ & $^{(c)}$ \\
\noalign{\smallskip}\hline
\end{tabular}
\end{table}

To ensure the dominant term in \equ{e:dynsys} is due to solar gravity only, we limit
the region of interest of our study in the space $(\beta,\gamma)$ as shown in light green
in Fig.~\ref{f:pars}. The dark shaded rectangle marks regions in $(\beta, \gamma)$ where central
gravity is less than twice the force due to solar radiation pressure (see top of
Fig.~\ref{f:forze}).

\paragraph{Choice of initial conditions.}
In this study we are interested in the behaviour of the orbital elements of the
charged particle in the vicinity of outer $p+q:p$ mean-motion resonances with
Jupiter, where the effect of radiation pressure on the mean-motion of the
dust particle, see \equ{e:FKep}, is taken into account. The value of the semi-major axis
of the dust particle at exact resonance is given by the relation 
\citep[equation 16 in][]{1994Icar..110..239B}:

\beqno
a_{p+q:p}=\left(\frac{p}{p+q}\right)^{2/3} a_1\left(1-\beta\right)^{1/3}
\eeqno

\noindent which follows from

\beqano
n^2a^3=\mu(1-\beta) \quad n_1^2a_1^3=\mu \ ,
\eeqano

\noindent and using the relation $n:n_1=p+q:p$. Here, $a_1$, $n_1$ denote the
semi-major axis and mean-motion of Jupiter. Let $\lambda=M+\omega+\Omega$,
$\lambda_1=M_1$ denote the orbital longitudes of the dust particle and Jupiter, respectively
(in agreement with the choice of reference frame such that the argument of perihelion and
longitude of the ascending node of Jupiter vanish). The resonant angular argument $s$ at
$p+q:p$ resonance is thus given by \citep[][]{1994CeMDA..60..225B, 1994Icar..110..239B}:

\beqno
q s=(p+q)\lambda_1 - p \lambda - q \bar\omega \ ,
\eeqno

\noindent where we used the notation $\bar\omega=\omega+\Omega$.

\section{Isolated influence of the interplanetary magnetic field}
\label{s:iso}

\begin{figure}
\begin{center}
\includegraphics[width=.40\linewidth]{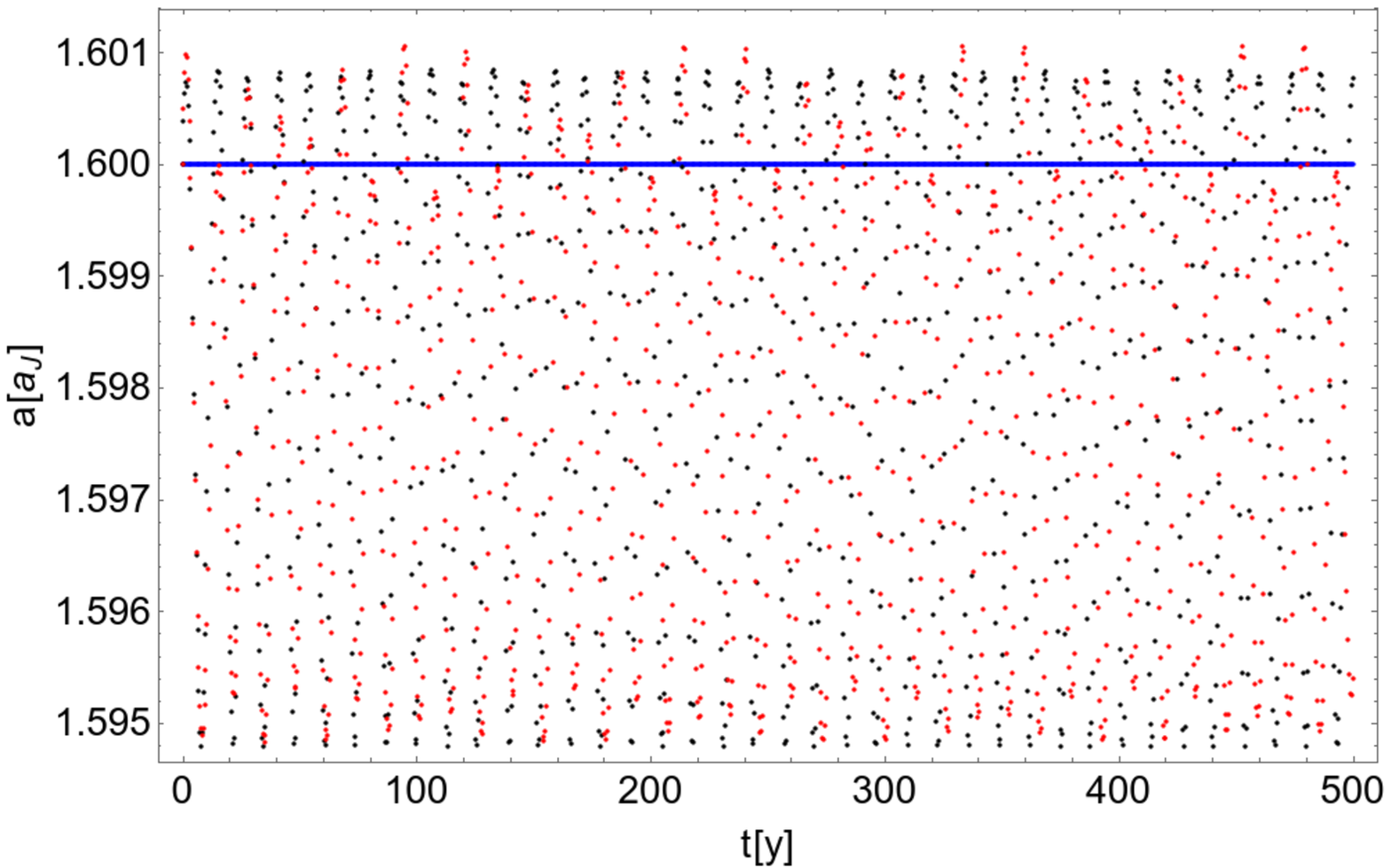}
\includegraphics[width=.40\linewidth]{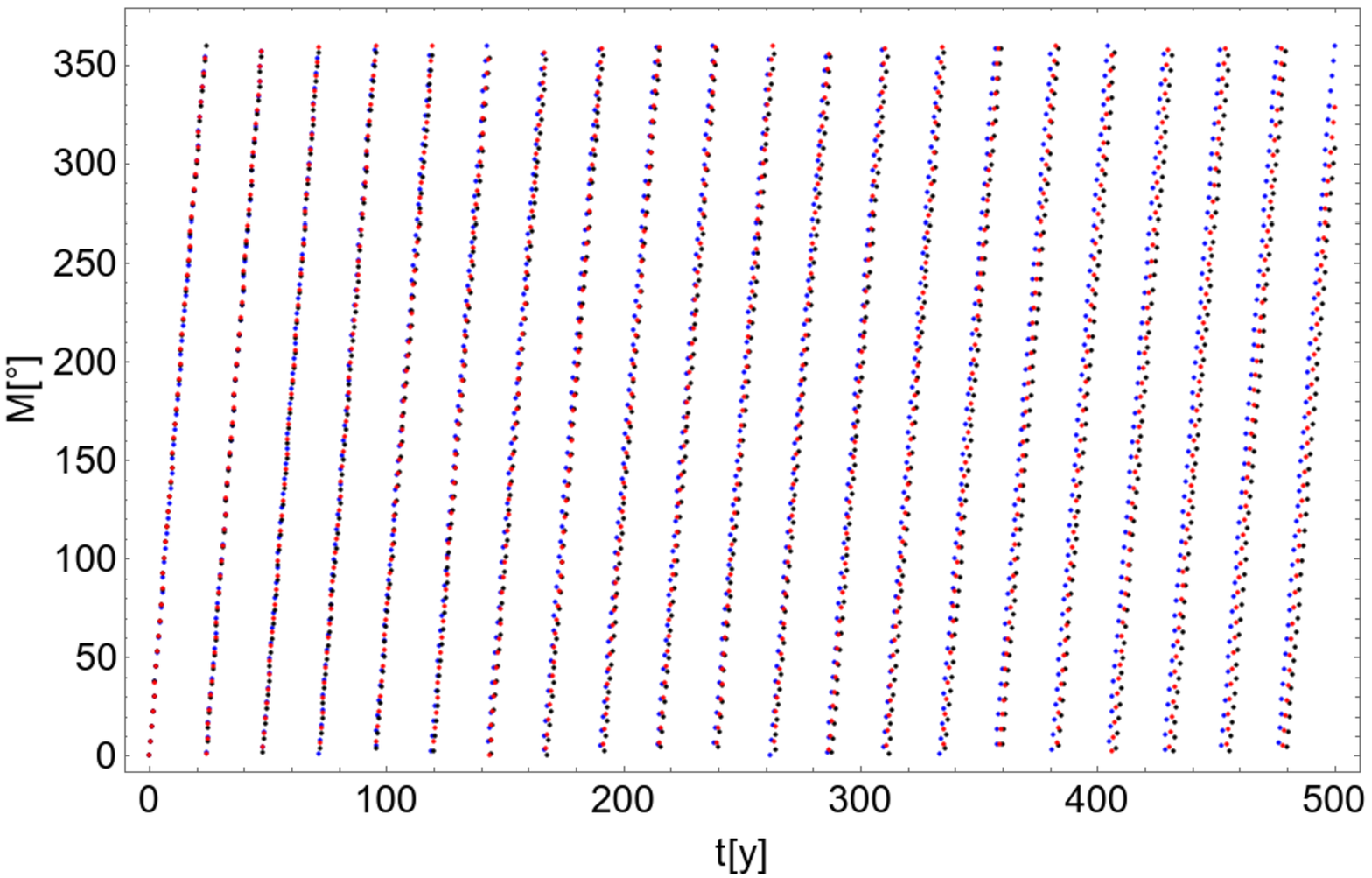} \\
\includegraphics[width=.40\linewidth]{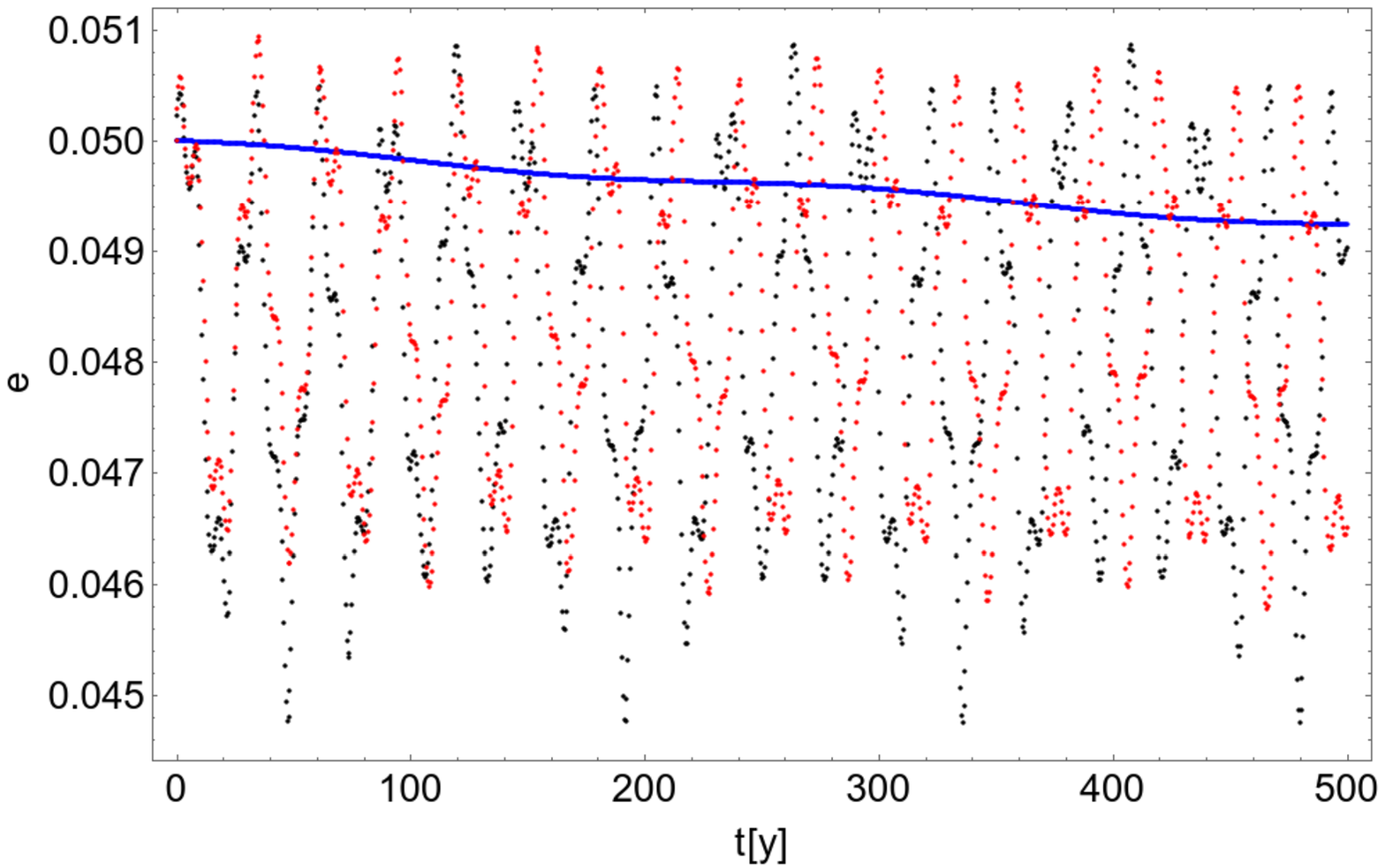}
\includegraphics[width=.40\linewidth]{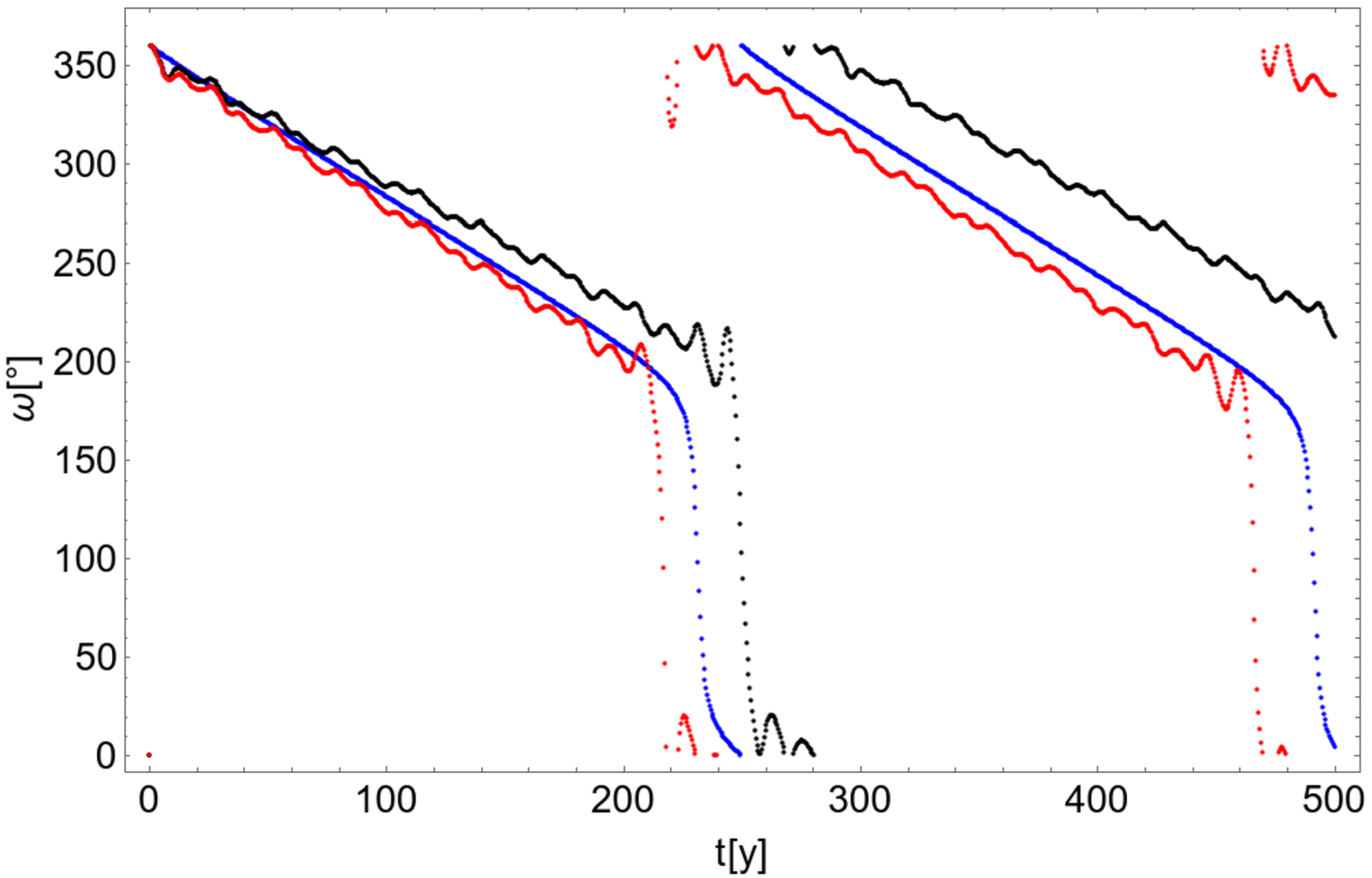} \\
\includegraphics[width=.40\linewidth]{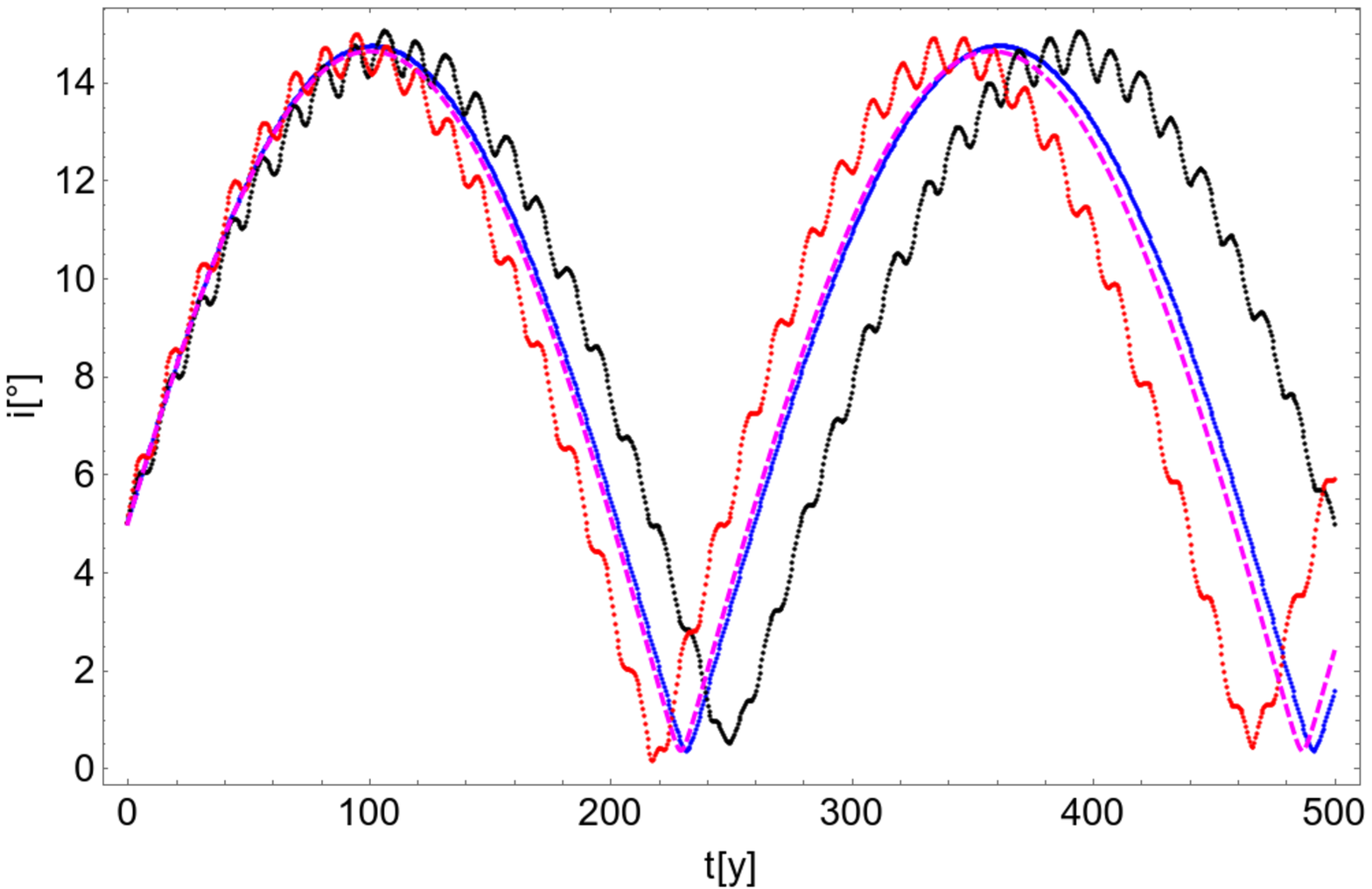}
\includegraphics[width=.40\linewidth]{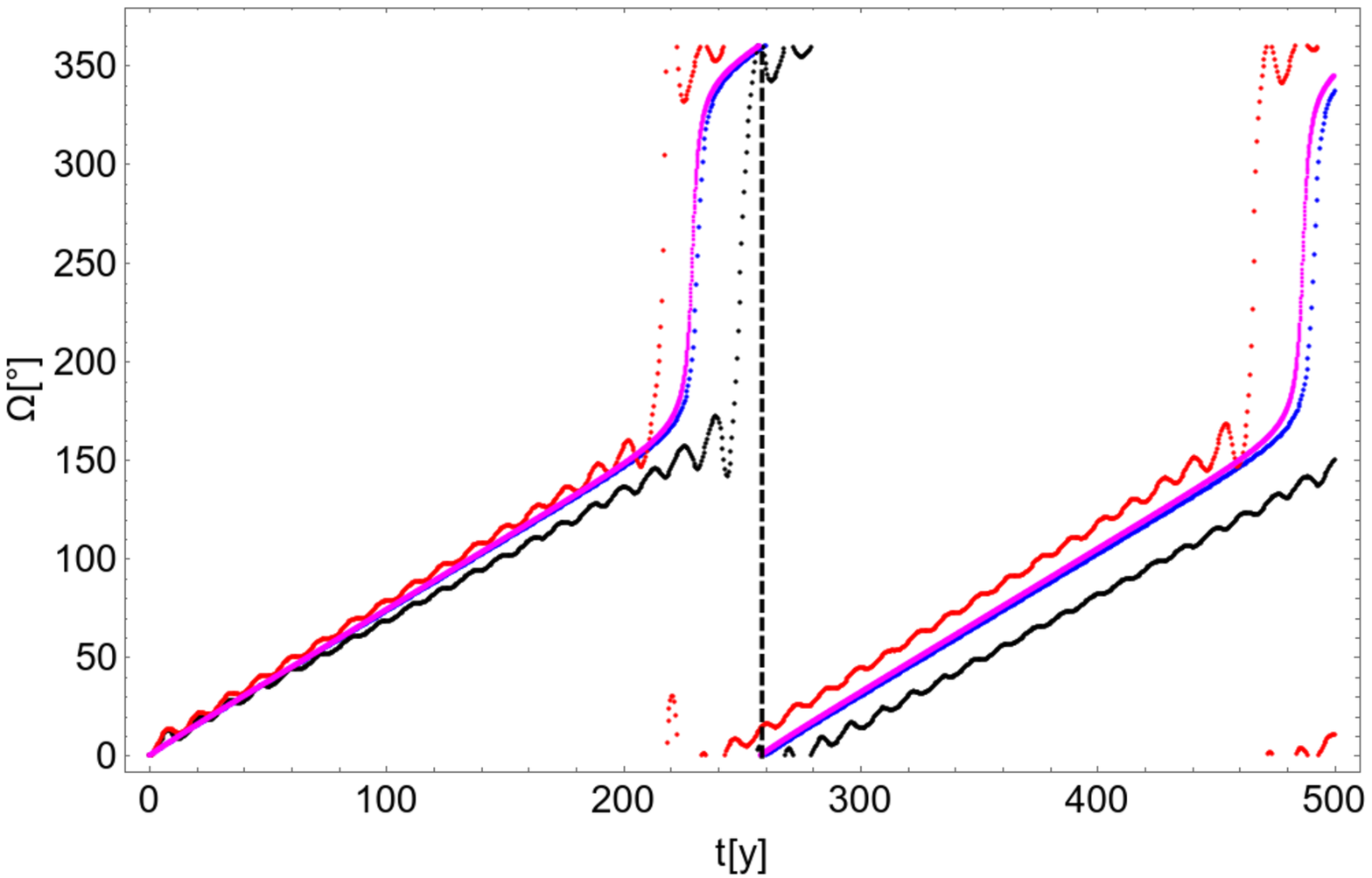}
\end{center}
\caption{Orbit for $a(0)=1.6 a_J$, $e(0)=0.05$, $i(0)=5^o$, and 
$\Omega(0)=\omega(0)=M(0)=0$. 
Comparison of Cartesian model (black), Gauss model (red), and averaged
Gauss model (blue). The dominant dynamics takes place in the $(\Omega,i)$-plane.
The magenta (dashed) line is the solution obtained from \equ{AF4},
the vertical dashed line in the lower right figure is calculated from the period of
\equ{AF5}.
}
\label{f:anaf}
\end{figure}

In this section we derive an analytical model that takes into account the
influence of the Keplerian part \eqref{e:FKep} and the Lorentz force
\eqref{e:cLF}. The aim of such a simplified model is to understand the orbital
evolution of a charged dust particle subject to the polarity change of the 
Sun's magnetic field in the long-term. The information provided by such a toy 
model is relevant in the study of the full model that also takes into account 
the influence of Jupiter, the Poynting-Robertson effect and solar wind drag. The 
analytical results are used in Section 4 to describe the complex phenomena 
arising in the neighbourhood of outer resonances with Jupiter.

The procedure for deriving this toy model is presented in detail in
\citet{2016ApJ...828...10L}.  Since this uses averaging theory, we underline
the fact that the analytical results will only be valid on the average over one
orbital period of the charged dust particle, and under the assumption that the
eccentricities are small, the inclinations have moderate values, and the
deviations of the magnetic axis from the angular momentum axis of the system
are also small.  Our approach differs from \citet{2016ApJ...828...10L} as
follows: 1) we allow the magnetic axis of the magnetic field to be misaligned
with the angular momentum vector of the Kepler problem. 2) In the present study
we also include a sign change function unlike our approach in
\citet{2016ApJ...828...10L}.

\noindent Our results in this section are based on the simplified dynamical problem:

\beq{simmod}
m\vec{\ddot{r}} = {\vec F}_{Kep} + {\vec F}_L \ ,
\eeq

\noindent with ${\vec F}_{Kep}$ given by \equ{e:FKep} and ${\vec F}_L$ given by
\equ{e:cLF}. The Lorentz force is not affected by the position of Jupiter.
The perturbation due to the planet is thus not taken into account in the
present discussion. Without loss of
generality, we set $\beta=0$ in \equ{e:FKep}.  First we derive Gauss' planetary
equations of motion \citep[see, e.g.][]{2012icm..book.....F} on the basis of
\equ{simmod}:

\beq{gauss}
\frac{d {\vec O}}{dt} = {\vec G}\left(\vec O\right) ,
\eeq

\noindent with $\vec O = \left(a,e,i,\omega, \Omega, M\right)$, and $\vec G$ is used
to denote the vector field that enters Gauss' planetary equations.  Starting from
\equ{e:cLF} we first express $r$, $\vec r$, $\dot{\vec r}$ in terms of series
expansions of the orbital elements by making use of Bessel functions
\citep[see, e.g.][]{dvolhobook}. The force ${\vec F}_L$ depends on the hyperbolic
tangent of the term ${\vec
r}\cdot{\vec g}_{\tilde z}/r$ that corresponds to $(x x_0 + y y_0 + z z_0)/r$
with $x_0=\sin\Omega_0\sin i_0$, $y_0=-\sin i_0 \cos \Omega_0$, $z_0=\cos i_0$, and $x$, $y$, $z$ are the components of $\vec{r}$ in the frame $\mathcal{G}_{Sun}$.
We make use of Lambert's continued fraction representation of the function
$\tanh$:  to zeroth order in eccentricity $e$, and for $x_0, y_0<<1$,
$z_0\sim1$ we find:

\beqa{af1}
\frac{\tanh\Bigl(\alpha\frac{{\vec r} \cdot {\vec g}_{\tilde z}}{r} \Bigr)}{\alpha}&\simeq&
\cos (\Omega ) \left(x_0 \cos (M+\omega ) + y_0 \cos (i) \sin (M+\omega )\right) \nonumber - \\
&& \sin (\Omega ) \left(x_0 \cos (i) \sin (M+\omega )-y_0 \cos (M+\omega )\right) \nonumber +  \\
&& z_0 \sin (i) \sin (M+\omega ) \ .
\eeqa

\noindent We note that \equ{af1} is valid for small deviations of the
magnetic axis from the inertial $z$-axis, for small eccentricities, and
moderate inclinations. Further calculations to obtain Gauss' planetary
equations in the form \equ{gauss} are straightforward
\citep[see, e.g.][]{2016ApJ...828...10L} and have been implemented using specialized
computer algebra routines in Wolfram Mathematica.  In the following exposition
we mainly address the secular contributions of the dynamics that are obtained
by averaging $\vec G$ over one orbital period of the dust grain expanded up to
first order in $x_0$, $y_0$, $1-z_0$, and vanishing $e$.  As numerical
simulations show \citep[e.g.][]{2006MNRAS.370.1876K}, the
mean effect of the Lorentz force mainly affects the orientation of the orbital
planes of the dust particles. In this section we confirm this result by
analysis of the averaged equations of motions. The averaged parts of
the force system $F_R$, $F_T$, $F_N$ in \equ{e:o2e} and \equ{gauss}
reduce to a non-vanishing normal component equal to:

\beqa{af2}
{\bar F}_N &\simeq& -\alpha\frac{3}{2}\frac{q}{m}B_0r_0^2\mu^{1/2} \frac{e}{a^{5/2}} \big\{
\cos (\omega ) \big[x_0 \cos (\Omega )+y_0 \sin (\Omega )\big] + \nonumber \\
&&\sin (\omega ) \big[\cos (i) \left(y_0 \cos (\Omega )-x_0 \sin (\Omega )\right)+z_0
   \sin (i)\big]
\big\} \ .
\eeqa

\noindent Since the averages of $\bar{F}_R$, $\bar{F}_T$ are zero we conclude that
the secular effects due to the interplanetary magnetic field mainly act on
inclination $i$ and on the ascending node $\Omega$, while the Lorentz force does not
alter the remaining Kepler elements on secular time scales. 
Ignoring the short-periodic effects,
we therefore fix the orbital elements $a$, $e$, and investigate the dynamics in the phase plane
($i$, $\Omega$). The components $i$ and $\Omega$ in Gauss' planetary 
equations of motions are given by \citep[see, e.g.][]{2012icm..book.....F}:

\beqa{AF3}
\frac{di}{dt} &=&
\frac{\cos\left(u\right)r F_N}{na^2\left(1-e^2\right)^{1/2}} , \ \nonumber \\
\frac{d\Omega}{dt} &=&
\frac{\sin\left(u\right)r F_N}{n a^2\left(1-e^2\right)^{1/2}
\sin\left(i\right)} \ .
\eeqa

\noindent To first order in $x_0$, $y_0$, $1-z_0$, and vanishing $e$, the
secular part that enters \equ{gauss}, \equ{AF3} reduces to:

\beqa{AF4}
\frac{di}{dt} &\simeq&
-\alpha\frac{q}{m}\frac{B_0}{2}\left(\frac{r_0}{a}\right)^2\bigg\{
\left[1 - \cos\left(i\right)z_0 \frac{\Omega_s}{n}\right]\times
\bigg(x_0\cos\left(\Omega\right)+y_0\sin\left(\Omega\right)\bigg)\bigg\}, \ \nonumber \\
\frac{d\Omega}{dt} &\simeq&-\alpha
\frac{q}{m}\frac{B_0}{2}\left(\frac{r_0}{a}\right)^2\bigg\{
\left[\cot\left(i\right)-z_0\frac{\cos\left(2i\right)}{\sin\left(i\right)}\frac{\Omega_s}{n}\right]\times \nonumber \\
&&\bigg(y_0\cos\left(\Omega\right)-x_0\sin\left(\Omega\right)\bigg)+\cos\left(i\right)\left(1-2z_0\right)\frac{\Omega_s}{n}+z_0\bigg\} \ .
\eeqa


\noindent We conclude that variations in inclinations and the ascending node
are proportional to the charge-to-mass ratio $q/m=\gamma [C/kg]$, the background magnetic
field strength $B_0$, and the square of the inverse distance of the dust grain
from the Sun $1/a^2$. Variations in inclinations are furthermore amplified by
the magnitudes of $x_0$ and $y_0$, proportional to a common proportionality factor
$1-\cos\left(i\right)z_0\Omega_s/n$. The dominant frequency
term in the second equation in \equ{AF4} depends on $\cos\left(i\right)$, the ratio
$\Omega_s/n$, and $z_0$, which is the cosine of $i_0$. We conclude that the 
overall effect of the perturbations due to the interplanetary magnetic field 
\citep[in absence of a normal magnetic field component, see][]{2016ApJ...828...10L}
on the orbital plane of a charged dust particle mainly depends on the actual 
value of $i_0$, inclination $i$, and the ratio $\Omega_s/n$. Let 

\beqno
\Gamma=\alpha \frac{q}{m} \frac{B_0}{2} \left(\frac{r_0}{a}\right)^2 \ .
\eeqno

\noindent In the limit $x_0,y_0\to0$ and 
$z_0\to1$, for fixed values of the semi-major axis $a$, and inclination $i$,
the solution for the ascending node becomes:

\beq{AF5}
\Omega(t) =
\left(\cos\left(i\right)\frac{\Omega_s}{n}-1\right)\Gamma t + \Omega(0) \ .
\eeq

\noindent Thus, for fixed $\Gamma$ the period of the ascending node decreases for
larger values of inclination $i$, and increases for larger values of $\Omega_s/n$.  
Next, we aim to solve \equ{AF4} in the limit $\Omega_s/n\to0$, which minimizes 
the period of the ascending node. Assuming small inclination $i$, we introduce the non-singular
and non-canonical variables:

\beqa{AF6}
p &=& \tan\left(i\right)\sin\left(\Omega\right) \nonumber \\
q &=& \tan\left(i\right)\cos\left(\Omega\right) \ ,
\eeqa

\noindent which transforms \equ{AF4} up to $O(p,q)$:

\beqa{AF6}
\frac{dp}{dt} = -\Gamma\left(y_0 + z_0q\right) \ ,
\frac{dq}{dt} = -\Gamma\left(x_0 - z_0p\right) \ ,
\eeqa

\noindent The linearised system has the solution:

\beqa{AF7}
p(t)\simeq\frac{\left(p_0 z_0-x_0\right) \cos \left(z_0 \Gamma t\right)-\left(q_0 z_0+y_0\right) \sin
   \left(z_0 \Gamma t\right)+x_0}{z_0} \ , \nonumber \\
q(t)\simeq\frac{\left(p_0 z_0-x_0\right) \sin \left(z_0 \Gamma t\right)+\left(q_0 z_0+y_0\right) \cos
   \left(z_0 \Gamma t\right)-y_0}{z_0} \ .
\eeqa

\begin{figure}
\begin{center}
\includegraphics[width=.55\linewidth]{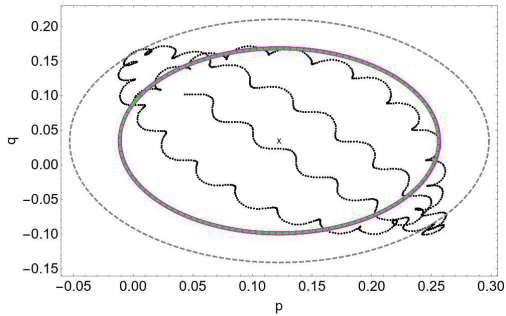}
\end{center}
\caption{Orbit of Fig.~\ref{f:anaf} projected to the $(p,q)$-plane. See text.}
\label{f:anaf2}
\end{figure}

\noindent From the solution it follows that the motion in the
$(p,q)$-plane takes place along a circle with radius $(x_0+y_0+q_0z_0)/z_0$
(with $q(0)=\tan\left(i(0)\right)$ and $p(0)=0$), centered around $(x_0,-y_0)$.
We note that the estimates are based on averaged and linearised equations that
are only valid for $x_0,y_0<<1$, $z_0\simeq1$, vanishing eccentricities, small
inclinations, and in the limit $\Omega_s/n\to0$. To demonstrate the validity of
the analytical results of the current section we show a specific orbit in
Fig.~\ref{f:anaf}.  The full, unaveraged solution obtained from
\equ{simmod} is shown in black, the unaveraged solution obtained from
\equ{gauss} is shown in red. The solution is well approximated with the
fully averaged model obtained from \equ{gauss}, and the numerical
solution of the reduced dynamical system \equ{AF4}, which is shown in magenta.
We note that the estimate of the period of $\Omega$ obtained from
\equ{AF5}, indicated by the dashed vertical line at the bottom right of
Fig.~\ref{f:anaf}, is in perfect agreement with the numerical solution of the
problem. However, the simplified solution given by \equ{AF5} is unable to
reproduce the complex evolution in time of the ascending node longitude
since it is lacking of higher order terms in its derivation. We
also show the same orbit projected on the phase-plane $(p,q)$ in
Fig.~\ref{f:anaf2}. The unaveraged solution based on \equ{simmod} is shown in
black, the averaged solution based on \equ{AF4} in magenta. The analytical
solution \equ{AF7} is indicated by the green line, and is confirmed by the
dashed circle with center and radius based on the considerations above.

We conclude that the dynamics in the $(i,\Omega)$-plane of charged dust in the heliosphere is
dominated by the deviation of the magnetic and orbital angular momentum axis and the rotation
period of the Sun. In our simplified
model the motion is regular and follows invariant curves in the $(p,q)$-plane. We note that
the rotation period in the $(p,q)$ plane, and thus the frequency of $\Omega$ strongly depend
on the choice for $\alpha$ that is present in the formulae due to the approximation 
of the $\tanh$ in \equ{e:cLF}. For the presentation of the results a reasonable choice of $\alpha$ has been
made to show the agreement between the averaged and unaveraged equations of motion. For
different values of $\alpha$ the frequencies may be over- or underestimated.

\section{Numerical study of the full problem}
\label{s:num}

In this section we perform several numerical experiments in order to highlight
the role of the Lorentz force on the long-term evolution of the orbital
elements of a charged dust grain located in the vicinity of outer mean-motion
resonances with Jupiter. By merging arguments given by analytical theories with
numerical investigations, in the following section we unveil a rich dynamical behavior
in the vicinity of resonance consisting of chaotic motions, trapped motions,
temporary captures, escapes from resonance, jumps, etc. Such phenomena are
highly influenced, or rather induced, by the complex interactions between the
Lorentz force and the other perturbations. For the purpose of exposition we
focus on the neighborhood of the 1:2 mean-motion resonance only. While
only specific test cases are shown to demonstrate various effects, many
numerical simulations have been performed to come to our conclusions.
Similar results are found close to the 1:3 and 2:3 mean-motion resonances. A
complete survey of the parameter and initial condition space for different
resonances will be the subject of a subsequent paper.

The dynamics of exterior mean-motion resonances has been studied in detail in
several works, such as \citet{1958AJ.....63..443M, 1993CeMDA..57..373S,
1994Icar..108...59L, 1994CeMDA..60..225B, 1994Icar..110..239B}, within both the
conservative planar circular restricted three body problem and the framework of
the dissipative model that includes, in addition, the influence of the
Poynting-Robertson effect. These studies have revealed a plethora of interesting
dynamical features of outer mean-motion resonances (as compared with interior
resonances) for example: the existence of asymmetric libration
regions, namely regions in which the resonant angle does not librate around
zero or $\pi$, temporary captures into resonances due to Poynting-Robertson
drag, the existence of a so-called universal eccentricity - that is an
eccentricity towards which all librational orbits will evolve regardless the
values of the drag coefficient or the planetary mass, etc.


Before describing the dynamical phenomena due to the
Lorentz force and the combined influence of the Lorentz force and
dissipative effects, let us recall first the structure of the phase portrait of
the 1:2 resonance. We disregard for the moment the influence of the
Lorentz force, Poynting-Robertson effect and solar wind drag - that is we
consider, as a starting point of discussion, the restricted three body problem
(the model that takes into account the effects of ${\vec F}_{Kep}$ and ${\vec
F}_{Jup}$) and we focus on the case of 1:2 resonance, located at
$a_{1:2}=2^{2/3} a_1\, (1-\beta)^{1/3}$ \citep[][]{1994Icar..110..239B}.

The dynamics of exterior mean-motion resonances, within the context of planar
restricted problem of three bodies, has been investigated in
\citet{1994CeMDA..60..225B} by using the Hamiltonian formalism. Starting
from a two-degrees-of-freedom non-autonomous Hamiltonian and assuming that the
system lies in the vicinity of an exterior mean-motion resonance, the problem
was reduced, by averaging the Hamiltonian with respect to the synodic period,
to a single-degree-of-freedom problem characterized by a Hamiltonian depending
on the action $J=L-G$, where $L$ and $G$ are the well-known Delaunay variables,
the critical angle $s$ and an external parameter $N$ that is a constant of
motion within the averaged problem \citep[see][]{1994CeMDA..60..225B}. For
the resonance considered here, $N$ has the expression \beq{e:N_Beauge}
N=\sqrt{\mu (1-\beta) a}\, \Bigl(2 \sqrt{1-e^2}-1\Bigr).  \eeq

\begin{figure}
\begin{center}
\includegraphics[width=0.49\linewidth]{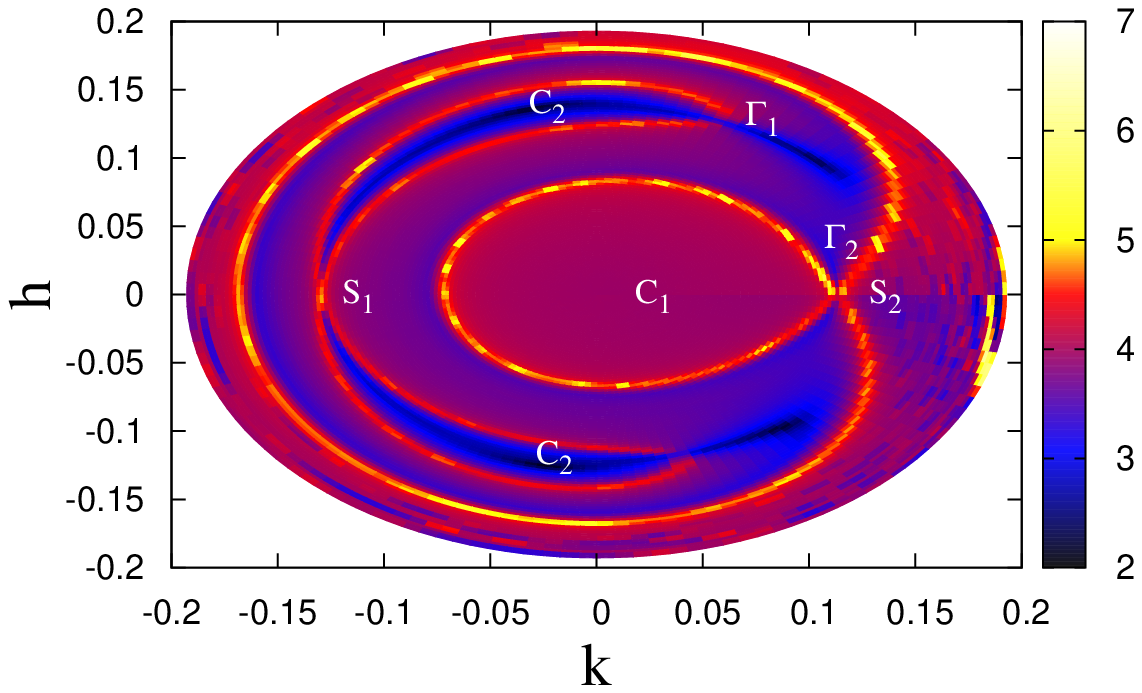}
\includegraphics[width=0.49\linewidth]{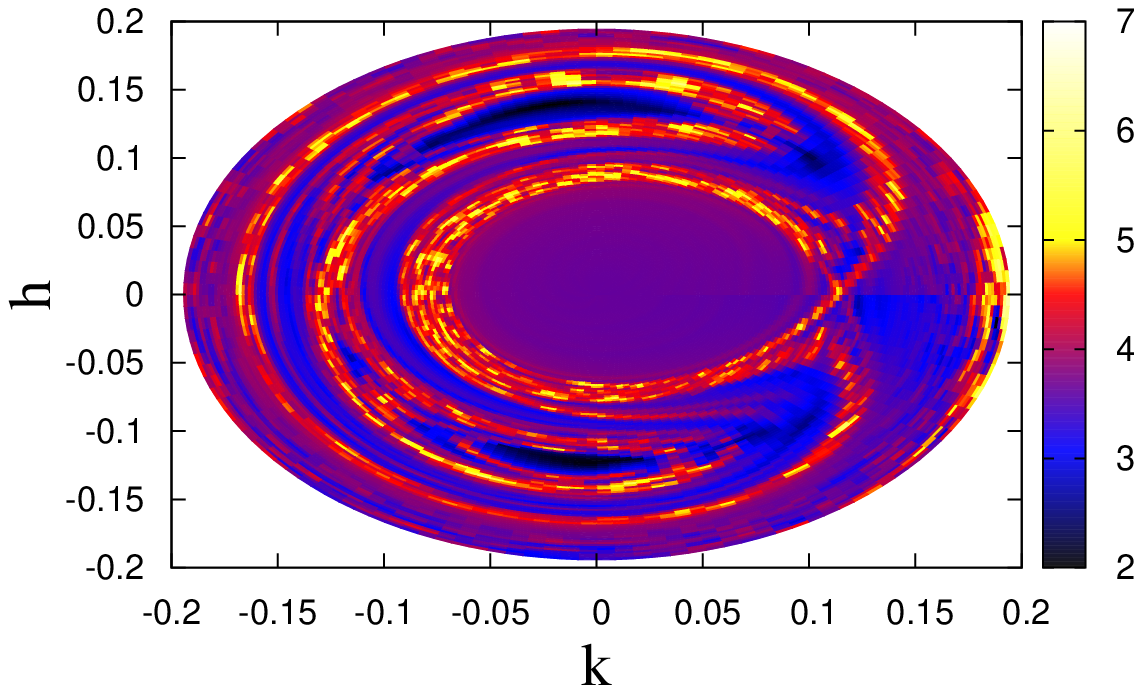}
\end{center}
\caption{FLI map for 1:2 resonance at $N=1.155$ with $i(0)=0^o.1$, $\omega(0)=0^o$,
$\Omega(0)=0^o$, $\beta=0.1$, $\gamma=0$ (left) and $\gamma=0.00227$
(right). Parameters correspond to particles with diameter $R=2\,[\mu m]$ and
$1\,[V]$ surface charge.
} \label{f:FLI}
\end{figure}

Left panel of Figure~\ref{f:FLI} shows the phase portrait of the 1:2 resonance,
within the context of the model defined by ${\vec F}_{Kep}$, ${\vec F}_{Jup}$
and the conservative part in \equ{e:swpr}. Results are shown in the non-canonical variables
$(k,h)=(e \cos (s), e \sin (s))$, for $N= 1.155$, in units of $\mu=1$, $a_1=1$
\citep[see,][]{1994CeMDA..60..225B}.  Using Cartesian equations of motion and
plotting the Fast Lyapunov Indicator, hereafter denoted by FLI
\citep[][]{FROESCHLE1997, GUZZO2002, GUZZO2013}, we numerically recover the
dynamical features of the 1:2 resonance, described in
\citet{1994CeMDA..60..225B} by using analytical arguments related to
single-degree-of-freedom Hamiltonians. Indeed, FLI reveals after a short
computational time the entire topological structure of the phase space. Namely:
two asymmetric libration regions (dark blue) with the centers $C_2$ situated at
$s\simeq \pm 95^o$, two separatrices $\Gamma_1$ and $\Gamma_2$ (yellow
lines) associated with the saddle points $S_1$ and $S_2$ respectively, an inner
circulation region (dark-red) with the center $C_1$, and the outer circulation
region. The separatrix $\Gamma_1$ surrounds the two asymmetric libration
regions, while $\Gamma_2$ splits the $(k,h)$ plane into the inner circulation
region, the libration zone (i.e. with orbits along which the critical angle $s$
does not take all values between $0^o$ and $360^o$) and the outer circulation
region. The libration zone includes the two asymmetric libration regions and
the region bounded by $\Gamma_1$ and $\Gamma_2$, which contain the orbits
exhibiting large-amplitude librations around the unstable point $S_1$.

Within the dynamical background described above, let us now consider the
effects induced by the Lorentz force~\eqref{e:cLF}. In this section, we
consider a fixed value of the parameter $\beta=0.1$, which correspond to a
sample object having the size $R=2 [\mu m]$, provided the density is
$\rho=2.8 [g/cm^3]$, and let the parameter $\gamma$ vary.  The right panel of
Figure~\ref{f:FLI} was obtained under the same conditions as the left panel,
but now the Lorentz force was "switched on".  We note that even if the
parameter $q/m=\gamma [C/kg]$ is small (namely the numerical value $\gamma=0.00227$,
corresponding to just $1\, [V]$ surface charge), the FLI map shows several yellow
regions that infer the onset of chaotic motions. Indeed, the color scale in
Figure~\ref{f:FLI} provides a measure of the FLI, which gives an indication of
the regular or chaotic dynamics: small values (i.e., dark colors) correspond
to regular motions, while larger values (i.e., red to yellow colors) denote
chaotic regions. The yellow regions suggest the appearance of chaotic motions,
that is some orbits might pass in an unpredictable way  from the libration
regime to the circulation regime, or from small-amplitude librations around one
of the centers $C_2$ to large-amplitude librations around the saddle point
$S_1$  and vice versa.

\begin{figure}
\begin{center}
\includegraphics[width=10cm,height=7cm]{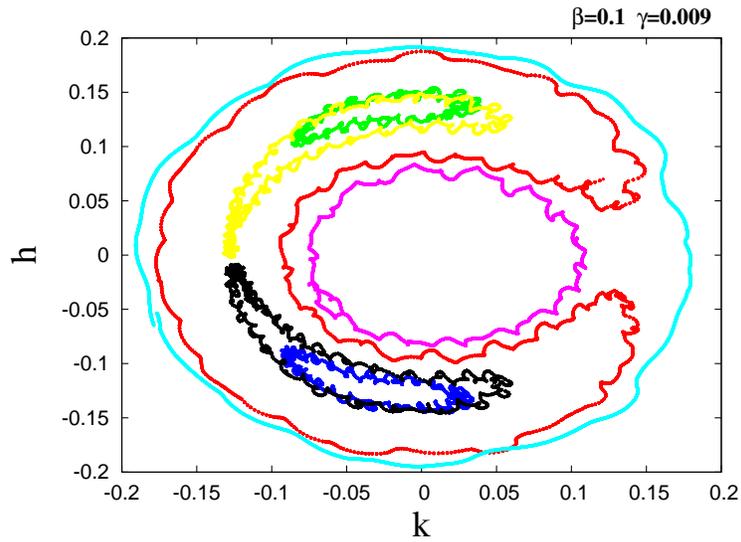}
\end{center}
\caption{Phase portrait for the 1:2 resonance, for $N=1.155$ in units of
$\mu=1$, $a_1=1$ (see the text), under the effect of the following
perturbations: Jupiter, the conservative part of the solar radiation pressure
and Lorentz force. Parameters $\beta=0.1$ and $\gamma=0.009$ correspond to
particles with diameter $R=2\, [\mu m]$ and $4\, [V]$ surface charge. The
osculating initial conditions of the seven orbits are $i(0)=0^o.1$,
$\omega(0)=0^o$, $\Omega(0)=0^o$ and: $e=0.14$, $s=30^o$ (red), $e=0.139$,
$s=80^o$ (green), $e=0.124$, $s=276^o$ (blue), $e=0.08$, $s=200^o$ (pink),
$e=0.19$, $s=200^o$ (light blue), $e=0.13$, $s=186^o$ (yellow), $e=0.13$,
$s=188^o$ (black).} \label{f:phase_space}
\end{figure}

Since chaos occurs for rather small values of $\gamma$, one may ask whether
librational motions are still possible for larger values of this parameter. The
question is also motivated by the analytical results described in the previous
section showing that the Lorentz force induces a large-amplitude variation of
inclination, up to $10$ degrees (see e.g. Figure~\ref{f:anaf}), which is
not present for $q=0$. The answer is provided by Figure
~\ref{f:phase_space}, obtained for the value of $\gamma$ that corresponds to
$4[V]$ surface charge. This Figure shows that in the $(k,h)$ plane all the
motions described within the framework of the planar restricted three body problem
are possible.  However, this plot was obtained by propagating seven orbits for
a relatively short period of time on the order of hundreds of years. Depending
on initial conditions and the parameter $\gamma$, on a longer time scale the
perturbations due to the Lorentz force lead to chaotic variations of the orbital
elements.

\begin{figure}
\begin{center}
\includegraphics[width=0.48\linewidth]{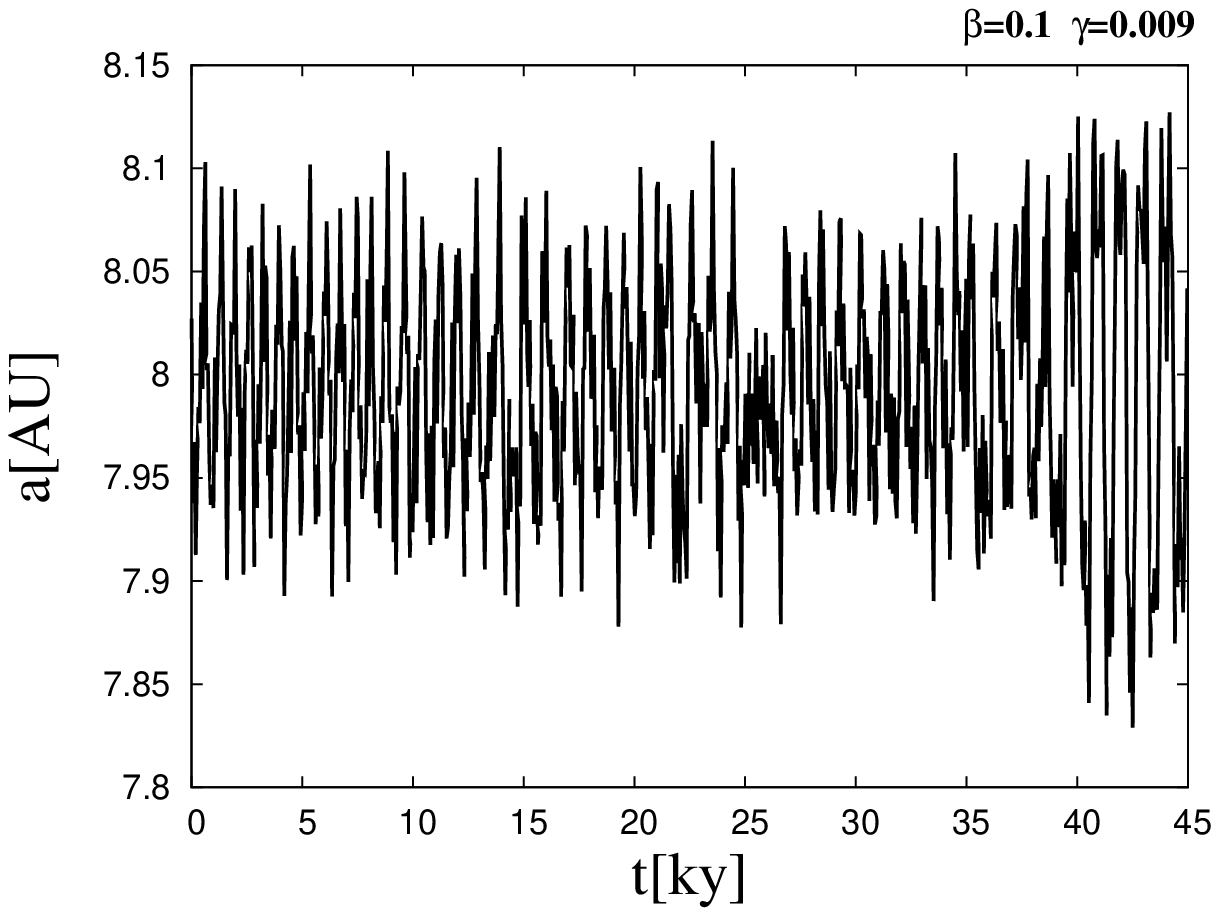}
\includegraphics[width=0.48\linewidth]{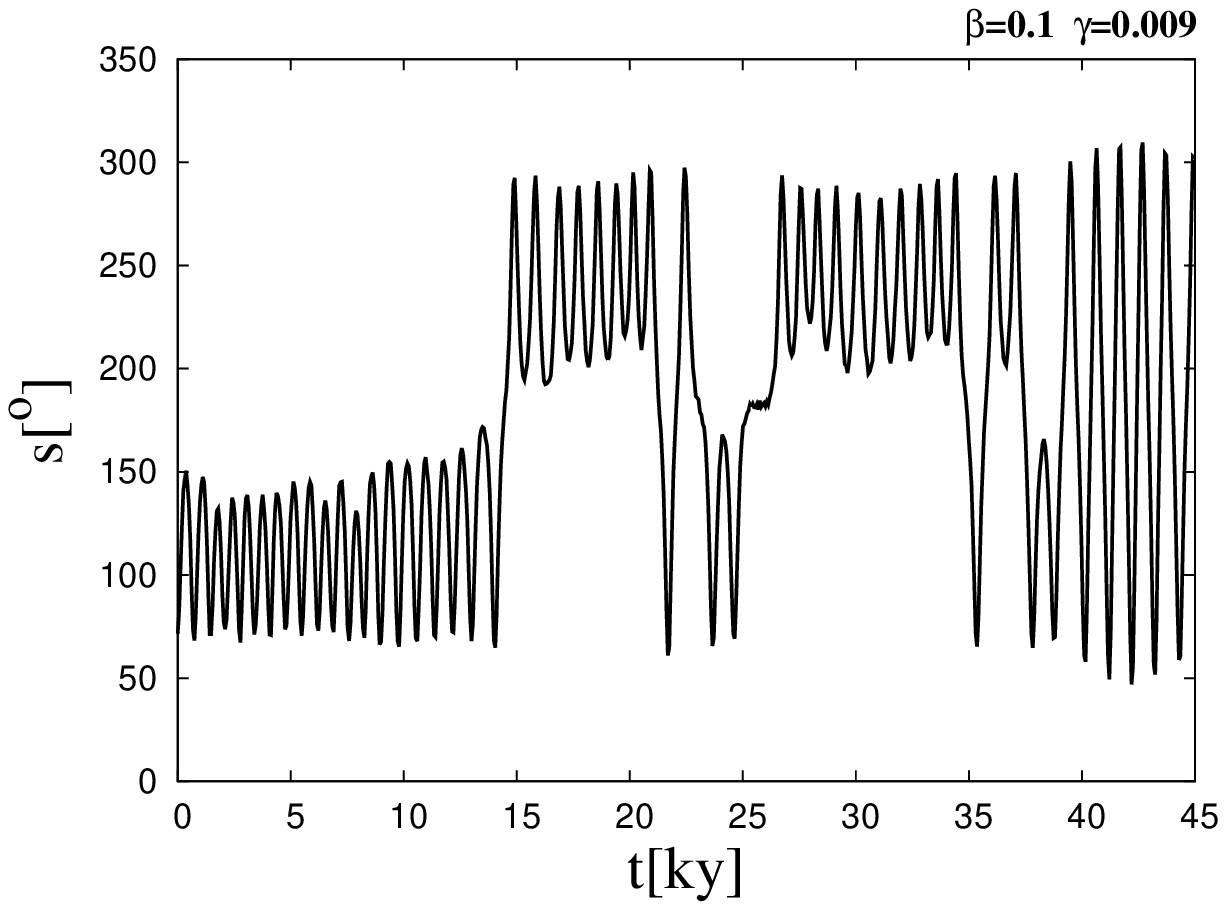}
\includegraphics[width=0.48\linewidth]{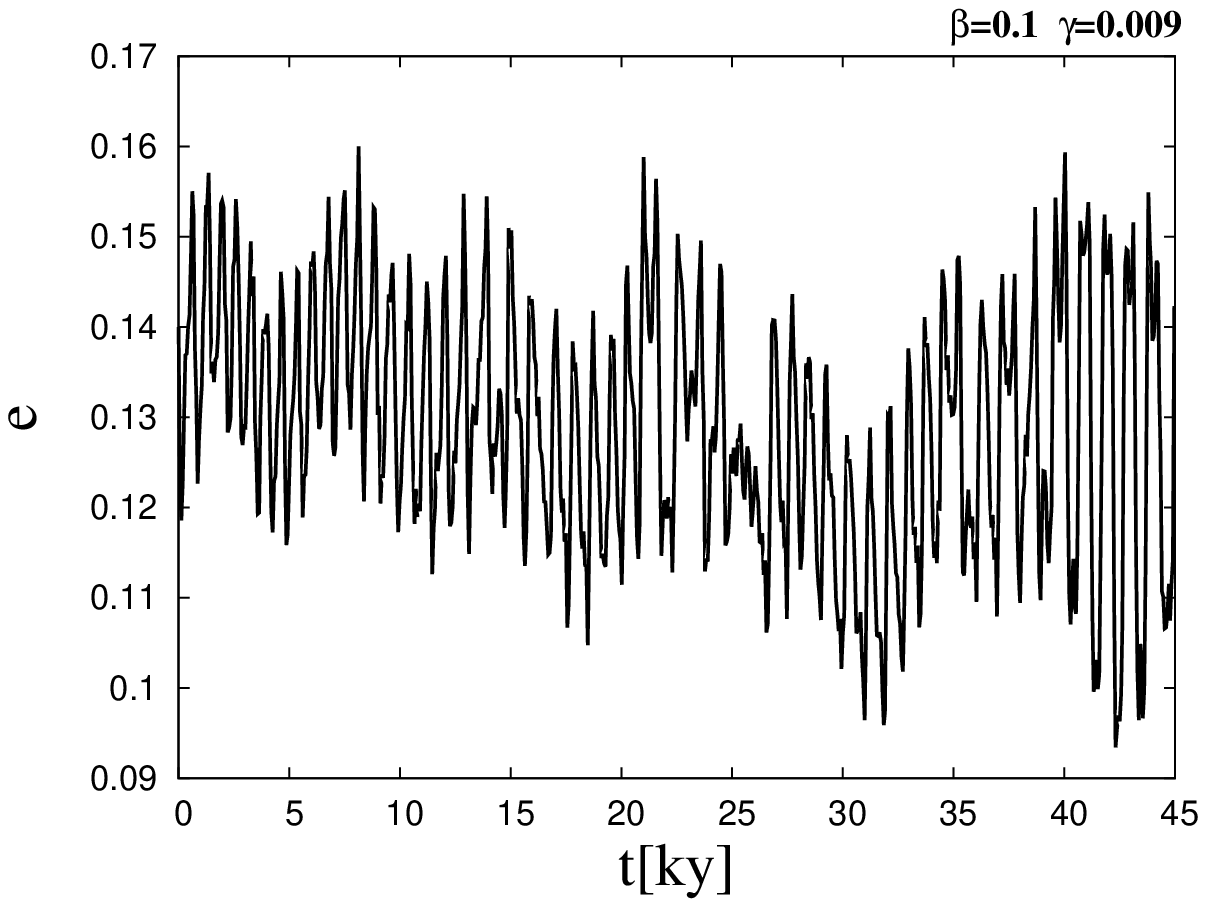}
\includegraphics[width=0.48\linewidth]{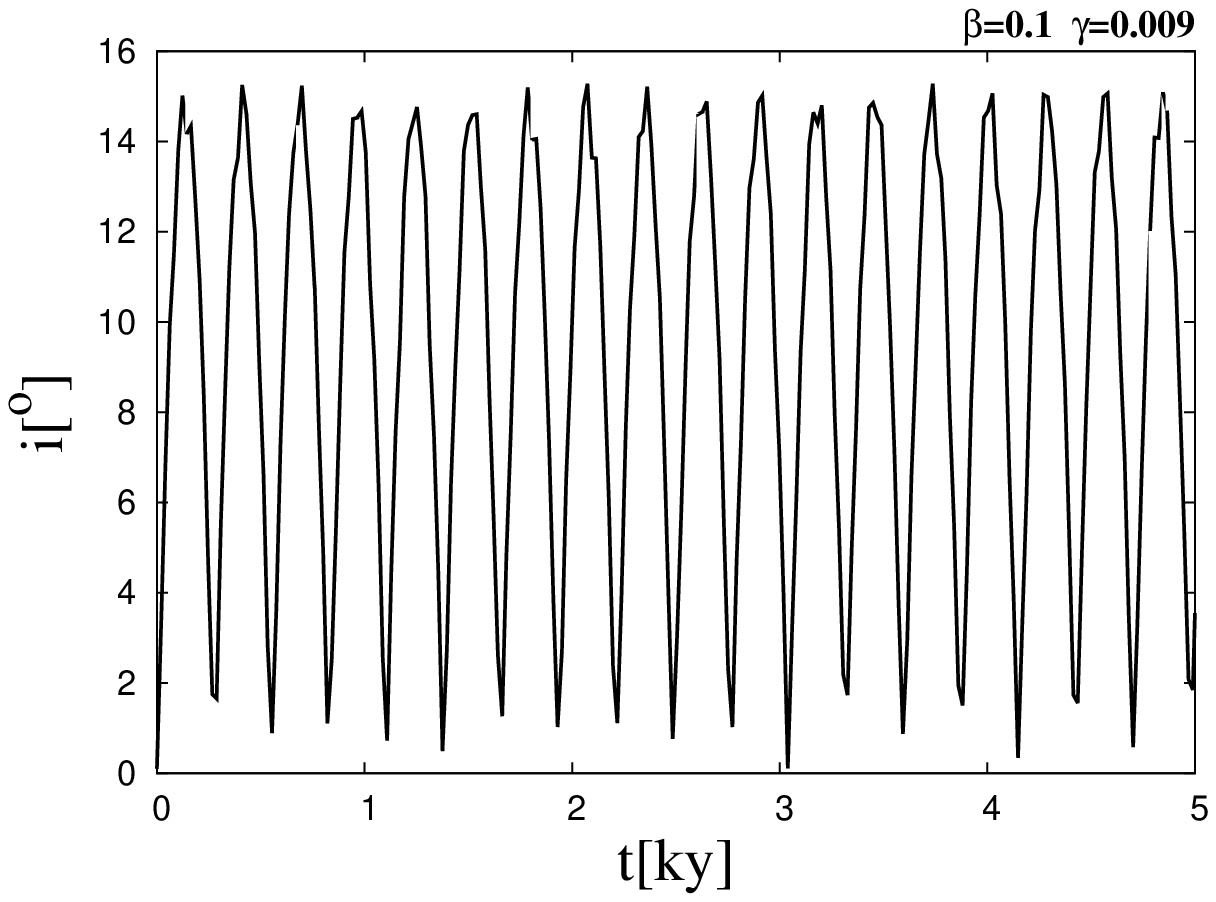}
\end{center}
\caption{Variation of semimajor axis $a$, resonant angle $s$, eccentricity $e$ and 
inclination $i$ for an orbit inside the 1:2 resonant regime. The parameters and osculating
initial conditions are: $N=1.155$, $\beta=0.1$,
$q/m=0.009 [C/kg]$, $e=0.14$, $s=72.5^o$, $i(0)=0^o.1$, $\omega(0)=0^o$ and
$\Omega(0)=0^o$. } \label{f:ex1}
\end{figure}

Figure~\ref{f:ex1} depicts a trapping trajectory into the resonance for more than
$45\, [ky]$. We remark that the initial conditions are close 
to the green orbit of Figure~\ref{f:phase_space}. Left panels show the evolution of
semi-major axis and eccentricity, while the right panels describe the behavior
of the resonant angle $s$ and the inclination $i$. The resonance does not
affect the evolution of inclination, which varies periodically due to the Lorentz
force (see the previous section); the right bottom plot of Figure~\ref{f:ex1}
shows the variation of inclination over 5$ \, [ky]$. One particular phenomenon
revealed by the top right panel of Figure~\ref{f:ex1} is the jump from one type
of librational motion to another and vice versa. Thus, as an effect of the Lorentz
force, this particular orbit exhibits short-amplitude librations around either
the center situated at $s\simeq 95^o$, or the stable point located at $s\simeq
-95^o$, and even large-amplitude librations around the saddle point $S_1$. Of
course, the jumps are unpredictable, and a small change in the initial
conditions leads to a different dynamical behavior.

\begin{figure}
\begin{center}
\includegraphics[width=0.48\linewidth]{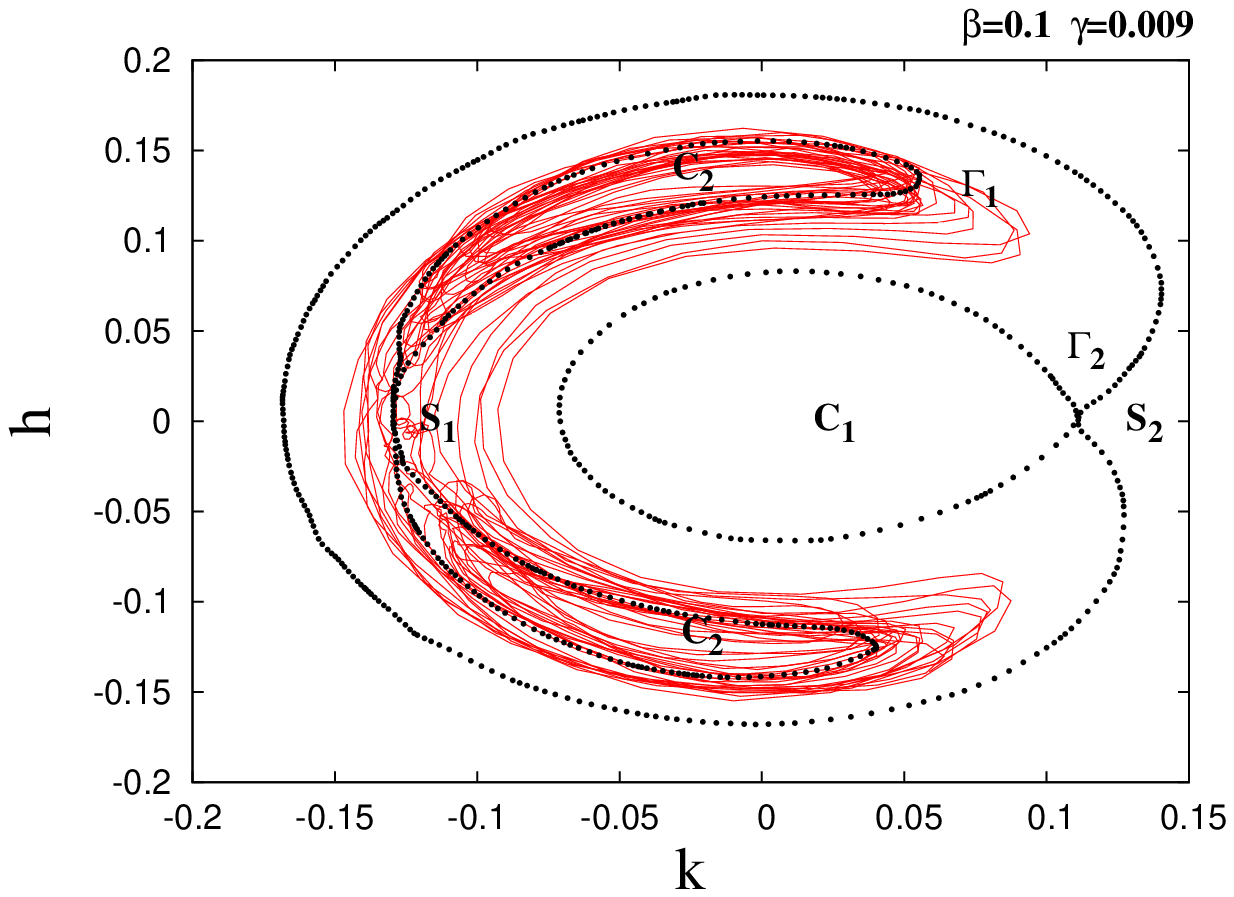}
\includegraphics[width=0.48\linewidth]{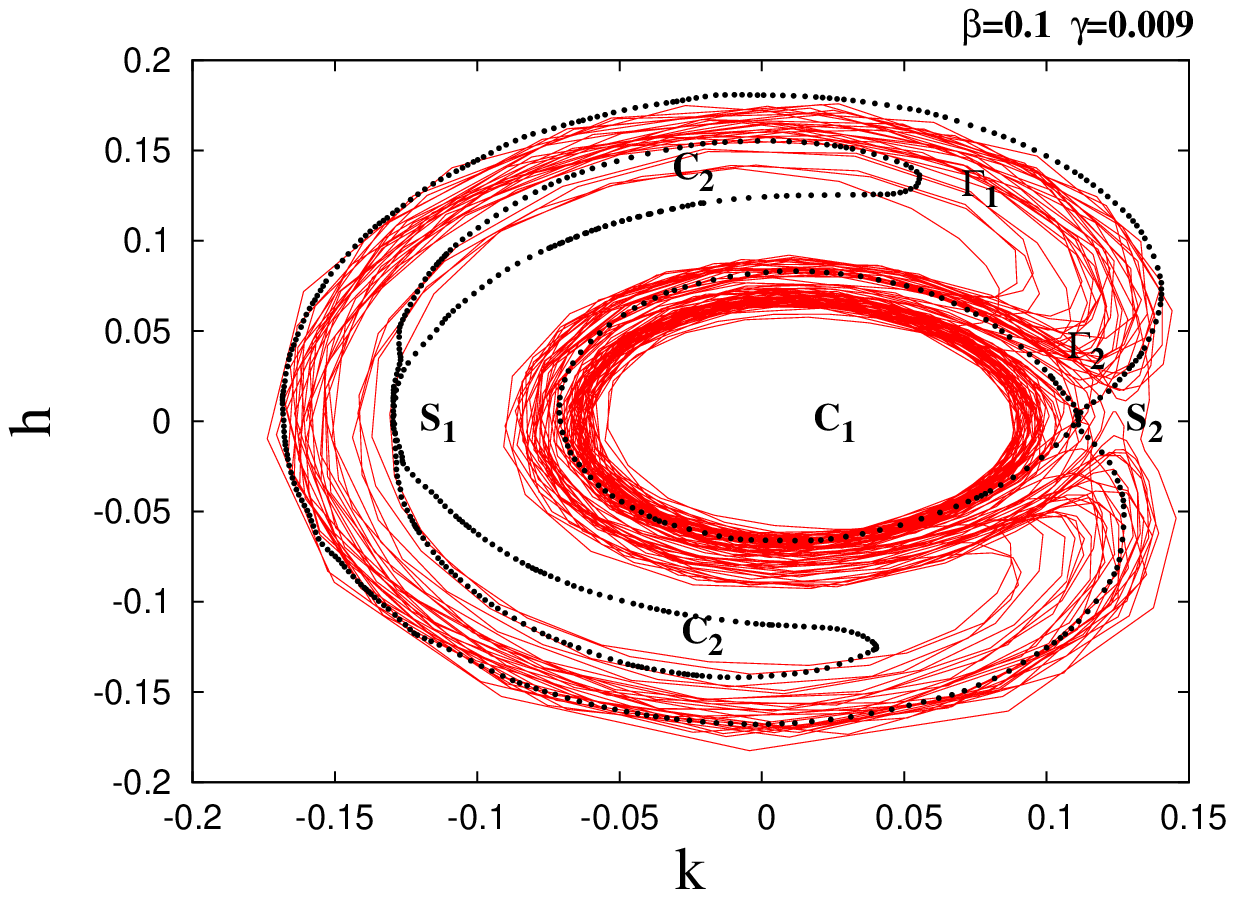}
\end{center}
\caption{Trapped motion into 1:2 resonance (left) and escape motion from it
(right), under the same effects as in Figure~\ref{f:ex1}. Left panel: variation
of the $(k,h)$ variables for the orbit shown in Figure~\ref{f:ex1}. Right
panel: the orbit characterized by the following parameters and initial
conditions: $N=1.155$ in units of $\mu=1$, $a_1=1$, $\beta=0.1$, $q/m=0.009
[C/kg]$, $e=0.16$, $s=180^o$, $i(0)=0^o.1$, $\omega(0)=0^o$ and
$\Omega(0)=0^o$. } \label{f:ex1_ex3}
\end{figure}

The left panel of Figure~\ref{f:ex1_ex3} shows the evolution in the $(k,h)$
plane of the orbit presented in Figure~\ref{f:ex1}.  To highlight various
dynamical behaviors, in particular the jumps from one libration region to
another,  on the background of Figure~\ref{f:ex1_ex3} we draw the separatrices
$\Gamma_1$ and $\Gamma_2$ from the planar restricted three body model.  In the
right panel of Figure~\ref{f:ex1_ex3} we present another dynamical behavior due
to the Lorentz force, that is the escape from the libration region of the 1:2
resonance. Indeed, this panel shows an orbit initially exhibiting
large-amplitude librations around the unstable point $S_1$, namely in the
libration region bounded by $\Gamma_1$ and $\Gamma_2$, but due to the
perturbations it crosses the separatrix and reaches into the inner circulation
region.

\begin{figure}
\begin{center}
\includegraphics[width=0.48\linewidth]{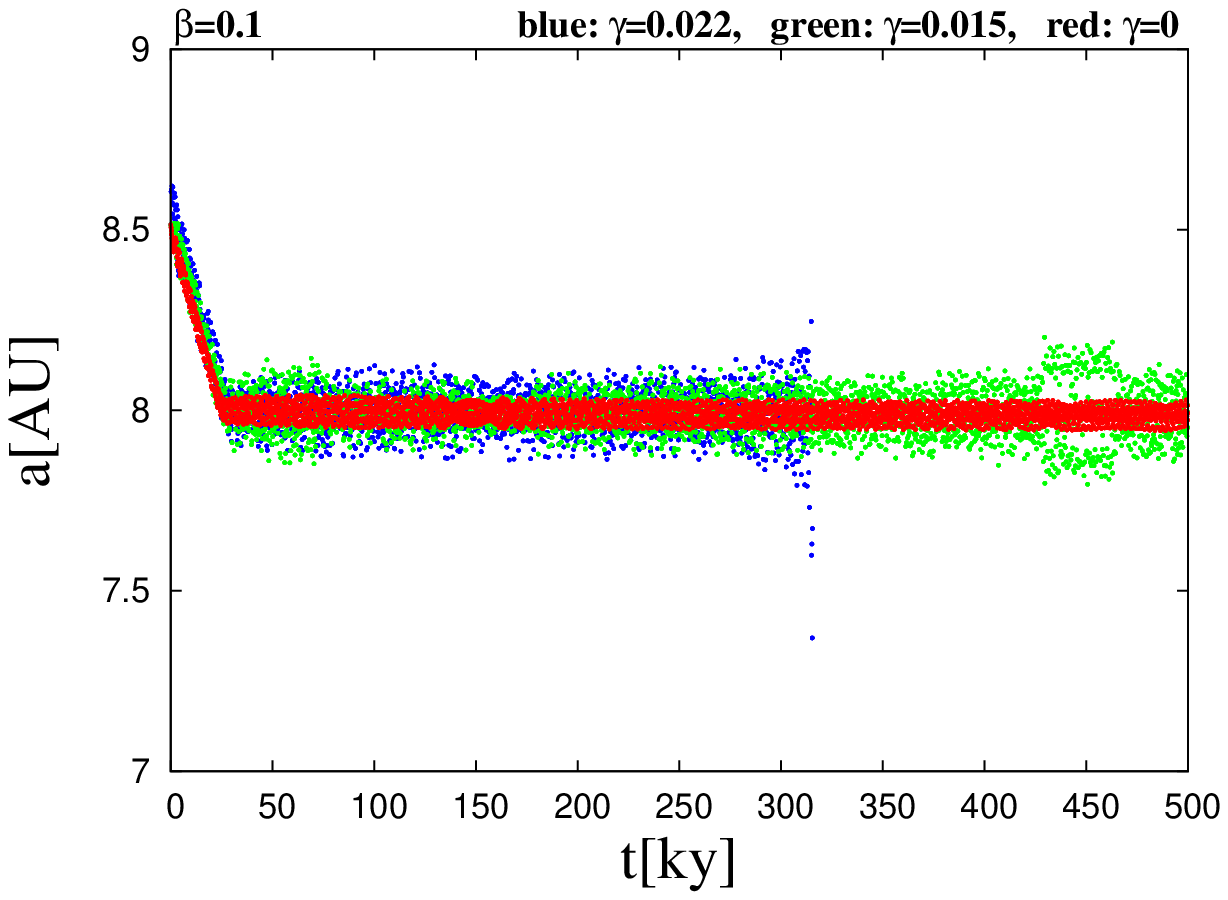}
\includegraphics[width=0.48\linewidth]{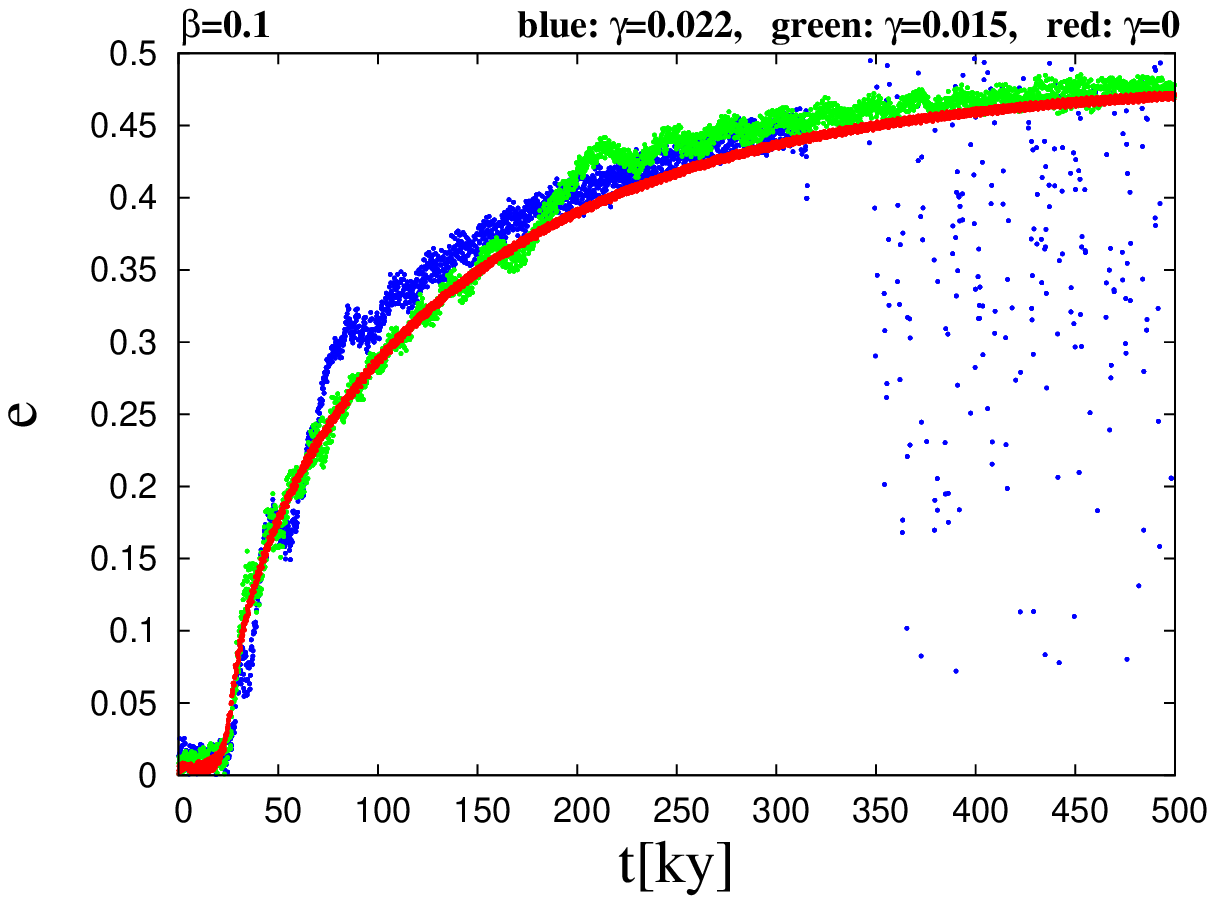}
\vspace{0.3cm}
\includegraphics[width=0.48\linewidth]{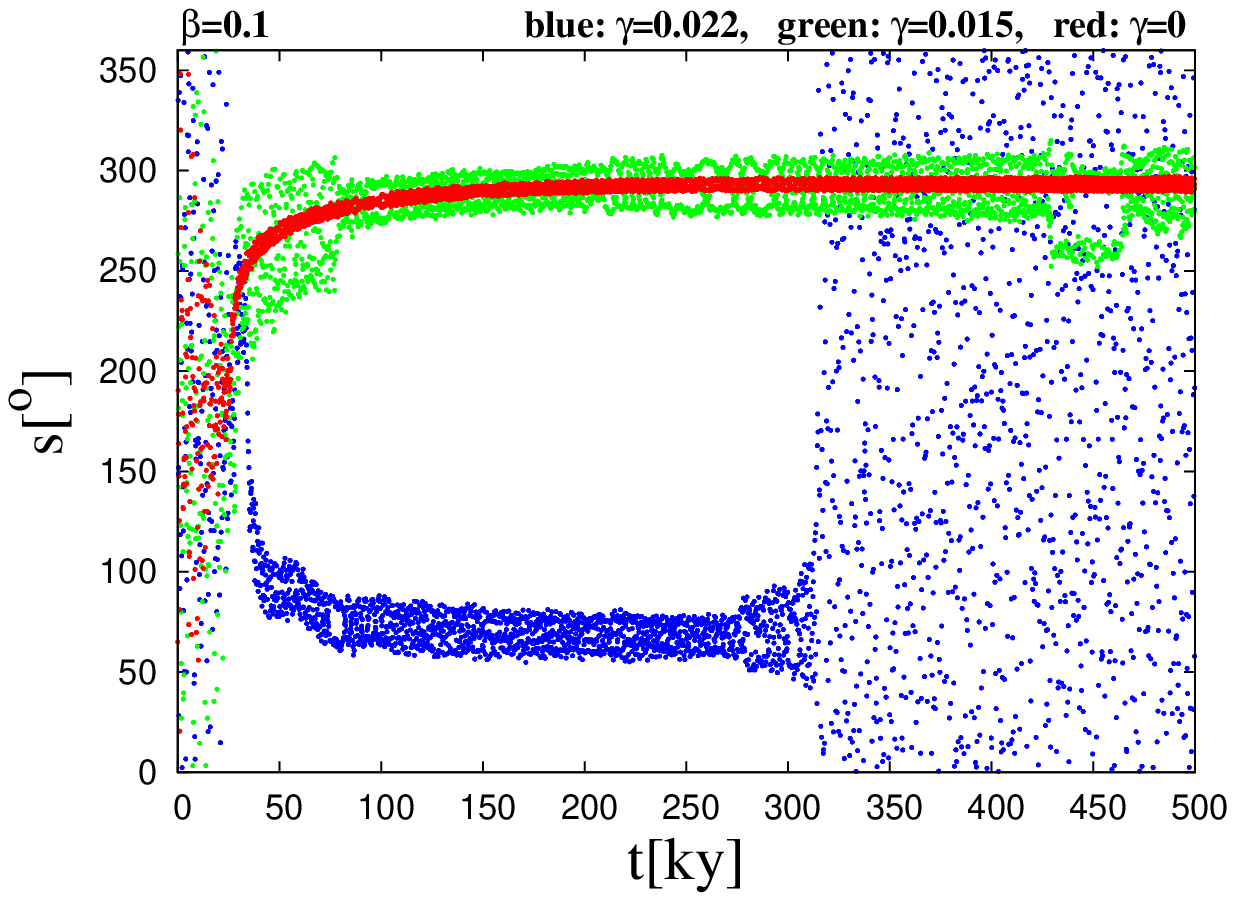}
\includegraphics[width=0.48\linewidth]{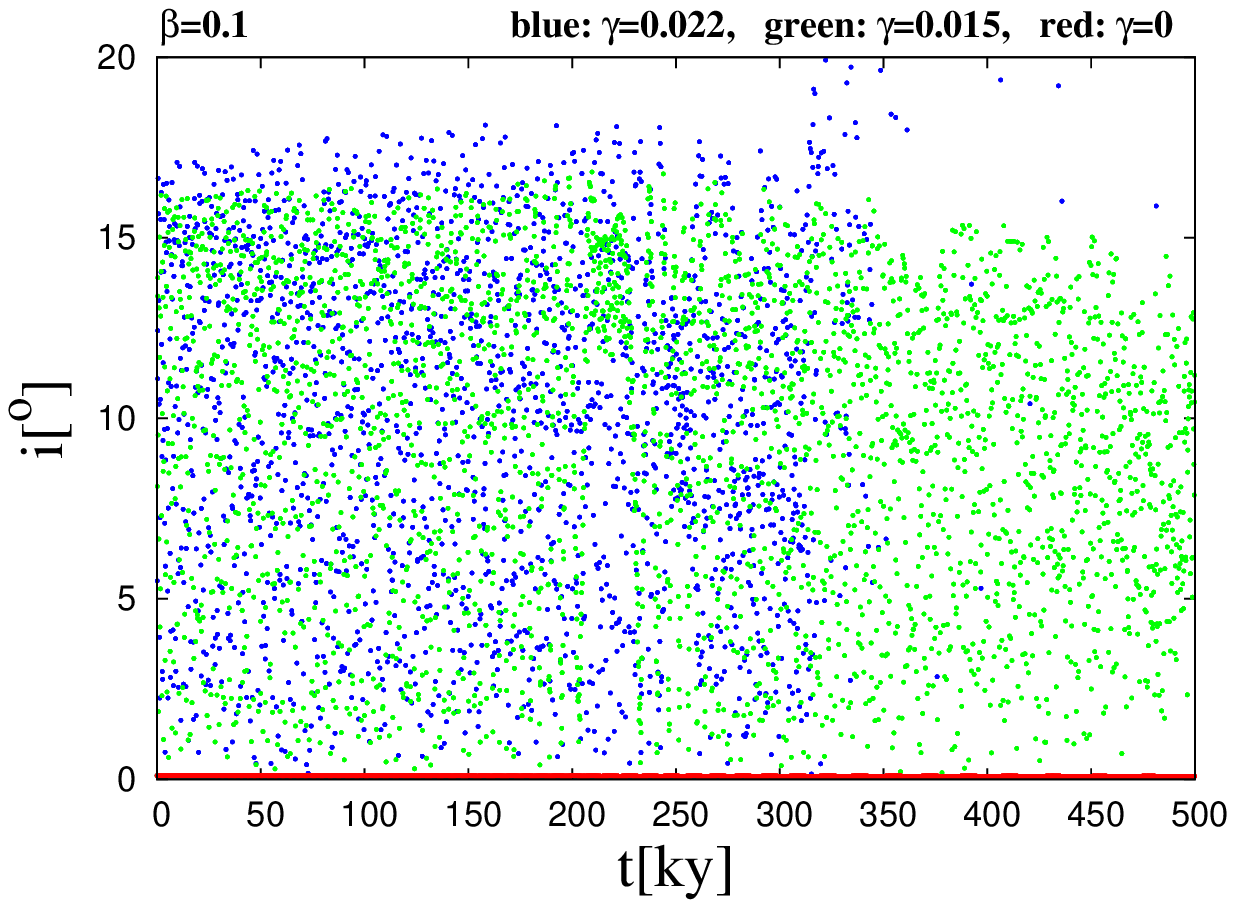}
\end{center}
\caption{Capture into the 1:2 resonance and escape from it. Variation of the
semimajor axis $a$, resonant angle $s$, eccentricity $e$ and inclination $i$
for three orbits. The orbits in green and blue are obtained under the
perturbing effects of the attraction of Jupiter, solar radiation pressure,
Poynting-Robertson drag and Lorentz force, while the orbit drawn in red was
obtained under the influence of Jupiter, solar radiation pressure and
Poynting-Robertson drag. The parameters and osculating initial conditions are:
$N=1.155$, $\beta=0.1$,  $e(0)=0.001$, $s(0)=0^o$, $i(0)=0^o.1$,
$\omega(0)=50^o$, $\Omega(0)=15^o$ and respectively $q/m=0.022 [C/kg]$ (blue),
$q/m=0.015 [C/kg]$ (green) and $q/m=0$ (red). The values $\gamma=0.022$
($\gamma=0.015$) correspond to about $9.7\,[V]$ ($6.6\, [V]$) surface charge.}
\label{f:ex4}
\end{figure}

The figures described above are obtained within the conservative framework (with all
forces except \equ{e:swpr}), and
provide an image of the dynamical behavior inside the resonance
(Figures~\ref{f:ex1} and \ref{f:ex1_ex3}). Let
us consider now the full dynamical model characterized by
equation~\eqref{e:dynsys} and analyze the orbital evolution of
non-resonant initial conditions.  Within the dissipative regime it is well
known \citep[][]{1993CeMDA..57..373S} that Poynting-Robertson effect leads
to temporary capture into exterior resonances, a dynamical mechanism that was
analyzed in~\citet{1994Icar..110..239B} by applying the adiabatic invariant
theory.  Moreover, it has been shown \citep[][]{1993CeMDA..57..373S,
1994Icar..108...59L, 1994Icar..110..239B} that there exists a so-called
universal eccentricity towards which all librational orbits will evolve
regardless of the values of the drag coefficient or the planetary mass. The
question we address here is the following: how does the Lorentz force \eqref{e:cLF}
influence the process of capture into resonance?  All numerical simulations
suggest that the Lorentz force does not influence the capture process itself; all
non-resonant initial conditions placed at a longer distance from the Sun than the location of resonance
evolve to a librational orbit. However, the
resulted trapped motions are shorter than in the absence of the Lorentz force.
Depending on the initial conditions and the value of $\gamma$, the orbits could be
trapped in resonance from $10^4$ to $10^6$ years.  Figure~\ref{f:ex4} reports
some results obtained by propagating three orbits having the same initial
conditions and the same parameters, but $q/m$ set to $\gamma=0$
(red), $\gamma=0.015$ (green) and $\gamma=0.022$ (blue). The values
$\gamma=0.015$ (green) and $\gamma=0.022$ (blue) correspond to large surface
charges of $6.6\, [V]$ and $9.7 \, [V]$, respectively.  These examples show
that even for large magnitudes of the Lorentz force the capture mechanism is
not affected. For the orbit in blue both the capture and escape phenomena are
clearly depicted by the bottom left panel of Figure~\ref{f:ex4}. The orbit in green
is captured into resonance for more than 450 years. We also note that orbits can be
captured in any of the two asymmetric libration regions. As a final remark, the
top right panel of Figure~\ref{f:ex4} provides a numerical evidence that also in this case there exists an universal eccentricity; eccentricities of the orbits trapped in
resonance evolve toward a value independently from the rest of the orbital
elements, in particular the inclination, and independent of the perturbations.

\begin{figure}
\begin{center}
\includegraphics[width=0.75\linewidth]{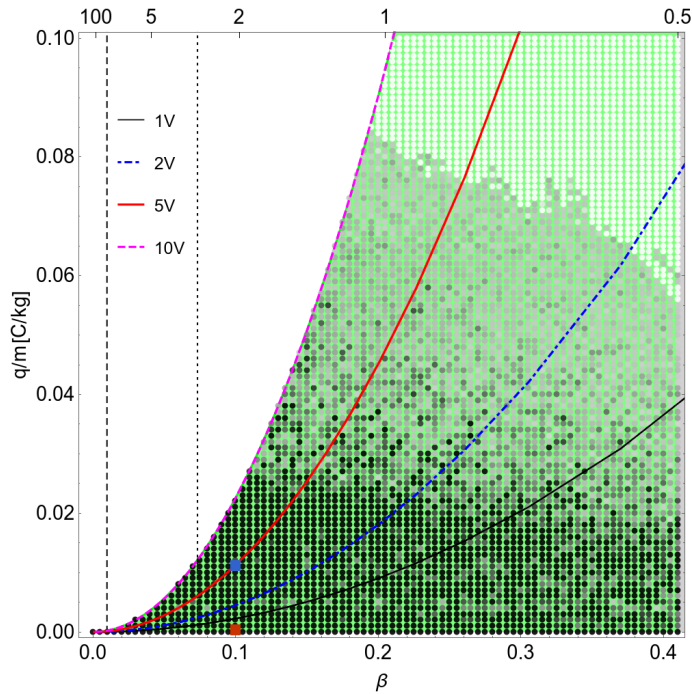}
\end{center}
\caption{Magnification of Fig.~\ref{f:pars}. Dots label the parameters $\beta,q/m=\gamma [C/kg]$
that have been used in our study. The color code indicates how long the dust grain
stays within the vicinity of the 1:2 resonance compared to the orbit with
$\gamma=0$ (black = 100\%, white = 0\%).}
\label{f:pars2}
\end{figure}

Finally, we perform numerical simulations in the full parameter space (green in
Fig.~\ref{f:pars}) on the grid $(\beta,\gamma)$ using discrete steps of $0.005$ in
$\beta$ and $0.001$ in $\gamma$. We start the dust grain on a circular orbit
within the orbital plane of Jupiter at the distance $a(0)=8.326 AU$  and $M(0)=0$.
We investigate the time for which the orbit of a charged grain stays in the
vicinity of the outer 1:2 resonance of the uncharged problem.  For this
reason we define the region $D_{1:2}$ as follows. The semi-major axis of the
uncharged dust grain shrinks until the particle is trapped at resonance. Let
$D_{1:2}$ be the interval $(0.95a_{min},1.05a_{max})$ within the time interval
$(0,100ky)$ of the orbit of the uncharged particle. Here, $a_{min}$ and
$a_{max}$ are the minimum and maximum values of $a(t)$ during the time of
integration equal to $T=100ky$ of the uncharged grain. For the charged particles we
determine the first moment in time, say $T_{s}$ when the semi-major axis of the
charged grain lies outside of $D_{1:2}$. We then calculate the ratio $T_{s}/T$
for each set of parameters $(\beta,\gamma)$, that is, the time for which the orbit
of the charged particle stays in the vicinity of the uncharged problem. We
note that the ratio does not imply that the orbit of the charged particle is
always trapped in resonance. However, it
serves as a rough measure for the stability in semi-major axis $a$. The results
are summarized in Fig.~\ref{f:pars2}.  Black dots indicate pairs of
$(\beta,\gamma)$ for which $T_{s}/T=1$ and white dots indicate $T_{s}/T=0$, while
intermediate cases are shown in gray. The figure clearly demonstrates that in
the majority of the cases, the time of temporary capture is decreased when
including the effects of the Lorentz force due to the interaction of charged particles
with the interplanetary magnetic field. 

\begin{figure}
\begin{center}
\includegraphics[width=0.45\linewidth]{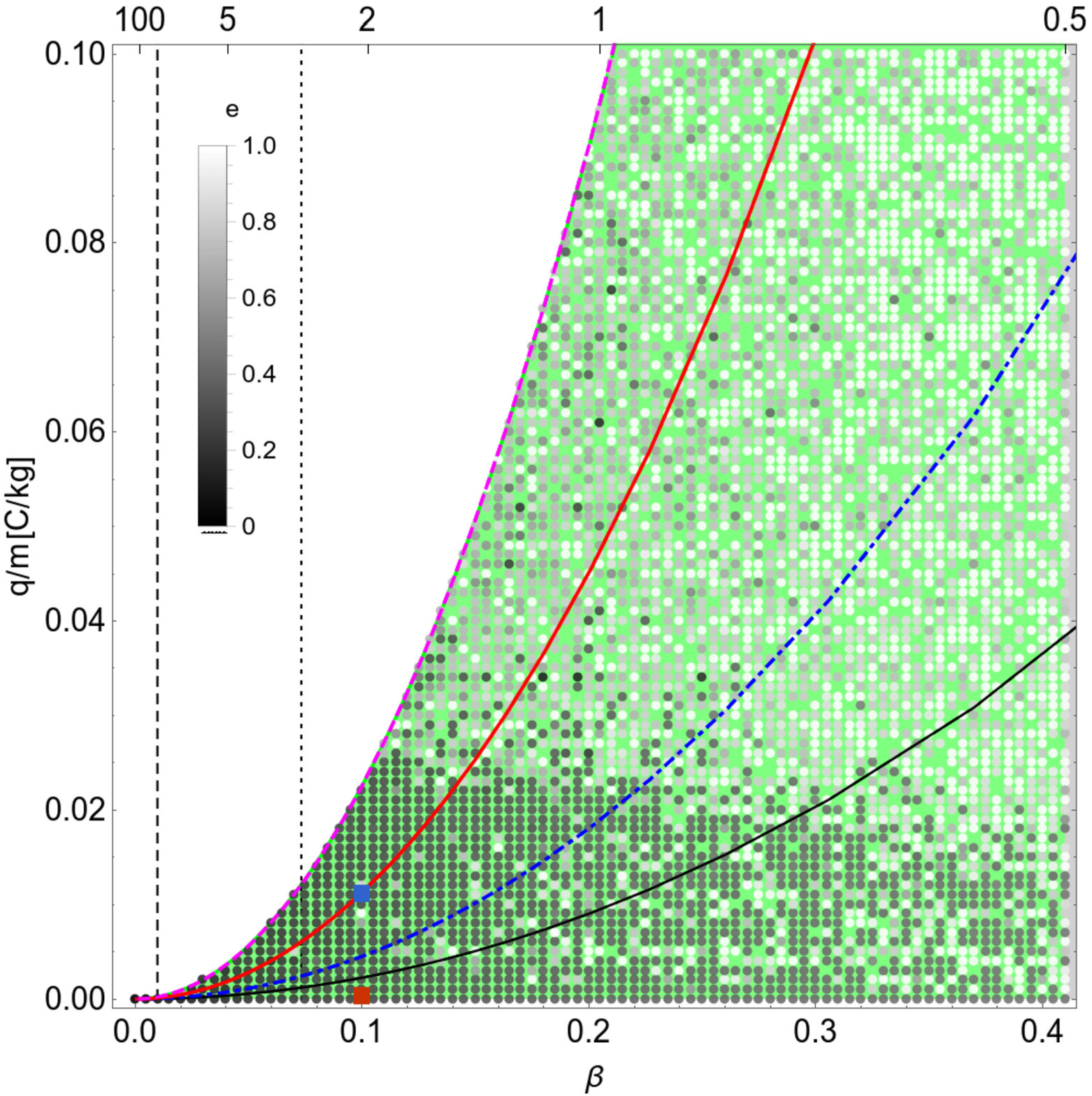}
\includegraphics[width=0.45\linewidth]{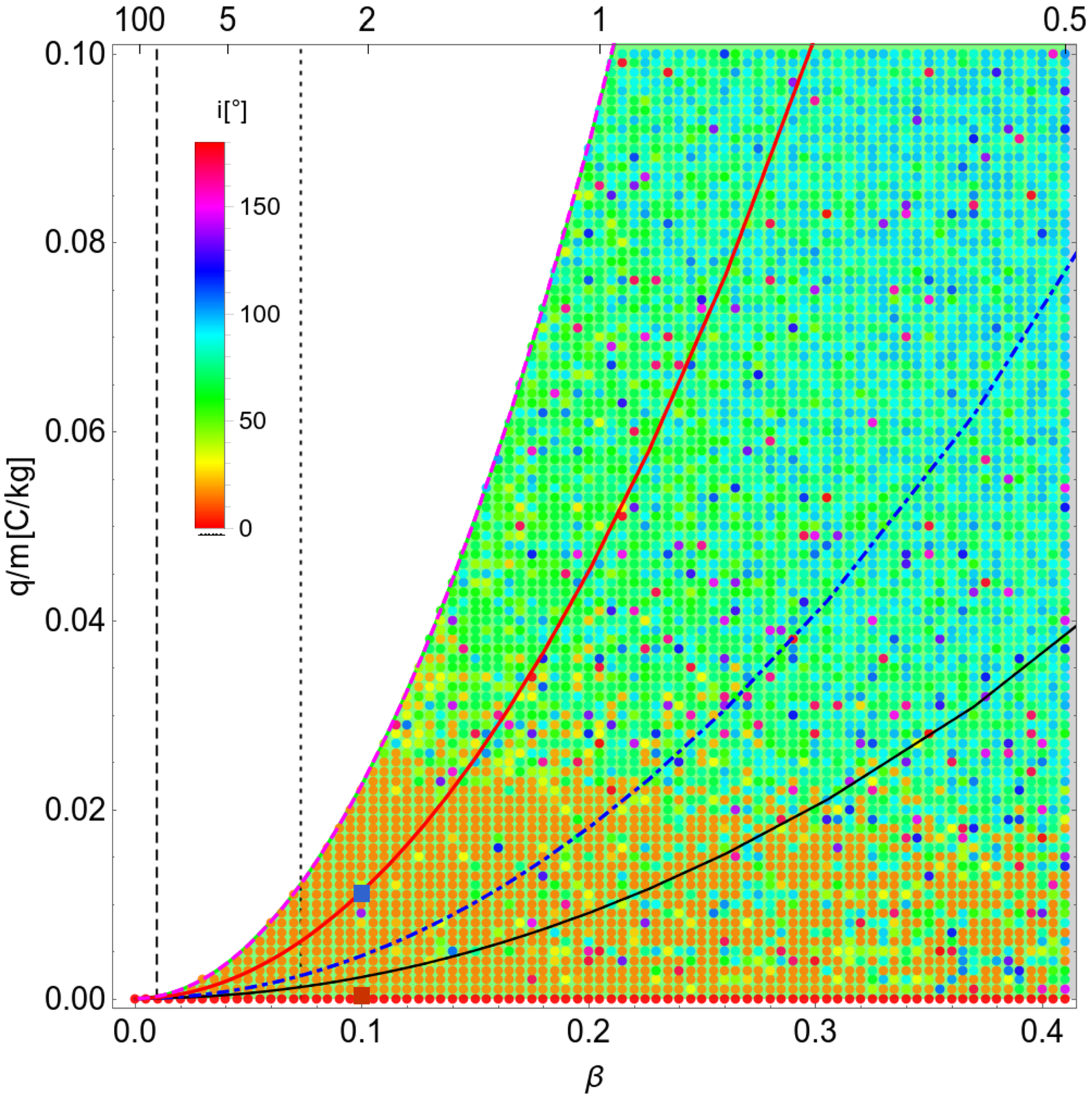}
\end{center}
\caption{Maximum values of $e$ (left) and $i$ (right)
in the parameter plane $\beta,q/m=\gamma [C/kg]$. Color code of lines correpsonds
to the color code of Fig.~\ref{f:pars2}. Within the orange (gray)
region variations in inclinations are less than $30^o$.}
\label{f:pars3}
\end{figure}

We note that in Fig.~\ref{f:ex4} inclinations vary with amplitudes
around $15^o$ for $\gamma\neq0$. The effect on the inclination stems from
the Lorentz force due to the interaction with the interplanetary magnetic field
(see model \equ{AF4} obtained in Section~\ref{s:iso}).  For $\gamma=0$, variations
in inclinations vanish, which can also be observed in the full model (see case
red in Fig.~\ref{f:ex4}).  To investigate the effect of $\gamma$ on the orbital
parameters, we look for the maximum values of $e$ and $i$ during their
evolution in time. The results are summarized in Figure~\ref{f:pars3}, where we
show $\max(e)$ and $\max(i)$ for the same
simulation data as in Fig.~\ref{f:pars2}. The black region in
Fig.~\ref{f:pars2} corresponds to the orange (dark gray) region in
Fig.~\ref{f:pars3}. Within this region, eccentricities are bound within ($0\leq
e \leq 0.7$) and inclinations stay within the interval ($0\leq i\leq 30^o$)
throughout the numerical integrations.  Outside this region eccentricities may
reach $e=1$ (white dots) and values beyond (blank) in Fig.~\ref{f:pars2},
while inclinations are found up to values about $i\simeq180^o$. As a direct
consequence, the possibility of collisions and close encounters of the dust
particle with planet Jupiter is increased, which will lead to reduced times in
temporary capture in comparison to the uncharged problem. We notice, that
the gray and white regions in Fig.~\ref{f:pars2} cannot be seen in 
Fig.~\ref{f:pars3}. The light gray region in Fig.~\ref{f:pars2} indicates the 
range of parameters $\beta$ and $\gamma$ for which the dust grains are captured -
about half the capture time of the uncharged dust grains with equivalent diameters. 
In this region, release from resonance occurs due to an increase of orbital parameters 
$e$ and $i$ somewhere in the middle of the orbital evolution, but the maximum 
values that have been calculated to produce Fig.~\ref{f:pars3} show values 
stemming from the second half of the orbital evolution of the dust grain orbits.
The same arguments hold true in the white region of Fig.~\ref{f:pars2}.
The mechanism for the reduction of temporary capture of dust grain orbits in 
mean-motion resonance with Jupiter can be explained by the enlargement of the 
chaotic regimes close to resonance (see Fig.~\ref{f:FLI}). As a result, the time of 
temporary capture in resonance is decreased - charged dust particles in the solar 
system are less stable than neutral ones.

\section{Summary of the results and discussion}
\label{s:sum}

In this work we investigated the effect of the Lorentz force due to the
heliospheric magnetic field on the orbits of charged, micron-sized dust grains
in the vicinity of outer mean-motion resonances with Jupiter with special focus on the
1:2 resonance.  We confirm previous results based on the assumption of
uncharged particles with some minor modifications. Our main results can be
summarized as follows:

\begin{itemize}

\item[1.]
The effect of the Lorentz force strongly affects the orbital planes of the dust
grains.  Therefore, the problem cannot be understood within the framework of
the planar problem.  We extend the problem to the spatial case, and find by
means of averaging theory the amplitudes of variations in orbital inclinations
of dust grains orbits. Moreover, the orbital planes experience a drift in
ascending node longitudes on secular time scales.  The effect on inclinations
is related to the angle between the angular momentum and magnetic axes of the
problem and the rotation rate of the Sun, see the discussion of
\equ{AF4} in Section~\ref{s:iso}.

\item[2.]
No major effect of the interplanetary magnetic field on the capture mechanism in
outer mean-motion resonances could be found. The capture process is dominated
by solar wind drag and the Poynting-Robertson effect like for uncharged
particles.  The transient time (before capture) turns out to be the same for
charged and uncharged particles, see the discussion of Figure~\ref{f:ex4} in
Section~\ref{s:num}.

\item[3.]
Resonant motion is present for charged dust grains like already
found for uncharged particles. However, the effect of the Lorentz
force induces additional perturbations that lead to complex dynamical phenomena. 
Most notably we find jumps of the resonant argument from one libration center to the 
other, and separatrix crossings (see the discussions in Section~\ref{s:num}).

\item[4.]
The presence of the interplanetary magnetic field affects the capture process.
The existence of the Lorentz force makes some dynamical processes faster, i.e.
escape from the resonant region (see, e.g. Figure~\ref{f:ex4}).

\end{itemize}

This work is the second in a series of papers devoted to the dynamics of
charged, micron-sized dust grains in the heliosphere \citep[see
also][]{2016ApJ...828...10L}. A detailed survey of the parameter space for
different outer mean-motion resonances is going to be finalized in the near
future and should include more sophisticated models of the interplanetary
magnetic field.

\vskip.1in

{\bf Acknowledgements}
This work is supported by the Austrian Science Fund (FWF)
with project number P-30542. CG was supported by a grant of the Romanian
National Authority for Scientific Research and Innovation, CNCS-UEFISCDI,
project number PN-III-P1-1.1-TE-2016-2314. We thank the members of the Space
Plasma Group at the Space Research Institute of the Austrian Academy of Science
for useful discussions about modelling the interplanetary magnetic field. \\

The authors declare that they have no conflict of interest.

\bibliographystyle{plainnat}
\bibliography{biblio}

\end{document}